\documentclass[12pt,english,american]{article}
\usepackage[T1]{fontenc}
\usepackage[latin9]{inputenc}
\usepackage{geometry}
\usepackage{color}
\usepackage{babel}
\usepackage{refstyle}
\usepackage{amsthm}
\usepackage{amstext}
\usepackage{amssymb}
\usepackage{mathrsfs}
\usepackage{amsmath}
\usepackage{cancel}  
\usepackage{graphics} 
\usepackage{graphicx}
\usepackage{yfonts}
\usepackage{soul}
\usepackage{tikz}
\usepackage[unicode=true,pdfusetitle,
 bookmarks=true,bookmarksnumbered=false,bookmarksopen=false,
 breaklinks=true,pdfborder={0 0 0},backref=false,colorlinks=true]
 {hyperref}
\hypersetup{
 linkcolor=blue,citecolor=blue, urlcolor=blue}

\makeatletter

\newcommand{\cI}{\mathcal{I}}

\def\qed{$\Box$\medskip}

\newcommand{\beq}{\begin{equation}}
\newcommand{\eeq}{\end{equation}}
\newcommand{\beqa}{\begin{eqnarray}}
\newcommand{\eeqa}{\end{eqnarray}}
\newcommand{\ben}{\begin{arabicenumerate}}
\newcommand{\een}{\end{arabicenumerate}}

\def\bel{\begin{lem} } 
\def\eel{\end{lem} }
\def\bet{\begin{thm}}
\def\eet{\end{thm}}
\def\bed{\begin{defn}}
\def\eed{\end{defn} }

\def\bec{\begin{cor}}
\def\eec{\end{cor}}
\def\ber{\begin{rem}}
\def\eer{\end{rem}}

\usepackage{enumitem}		
\theoremstyle{plain}
\newtheorem{thm}{\protect\theoremname}[section]
\theoremstyle{definition}
\newtheorem{defn}[thm]{\protect\definitionname}
\theoremstyle{plain}

\theoremstyle{plain}
\theoremstyle{remark}
\newtheorem{rem}[thm]{\protect\remarkname}
\theoremstyle{plain}
\newtheorem{lem}[thm]{\protect\lemmaname}
\theoremstyle{plain}
\newtheorem{cor}[thm]{\protect\corollaryname}

\usepackage{txfonts,refstyle,xcolor}

\newref{con}{name = Conjecture\ }
\newref{prop}{name = Proposition\ }
\newref{def}{name = Definition\ }
\newref{sec}{name = Section\ }
\newref{sub}{name = Section\ }
\newref{thm}{name = Theorem\ }
\newref{lem}{name = Lemma\ }
\newref{cor}{name = Corollary\ }
\newref{fig}{name = Figure\ }
\newref{rem}{name =  Remark\ }

\usepackage{bbm}
\newcommand{\charf}{\mathbbm{1}}

\usepackage[all]{xy}

\newcommand{\xyR}[1]{%
     \makeatletter
     \xydef@\xymatrixrowsep@{#1}
     \makeatother
}

\newcommand{\xyC}[1]{%
     \makeatletter
     \xydef@\xymatrixcolsep@{#1}
     \makeatother
}

\newcommand{\ncol}[1]{\color{normalcolor}}

\makeatother

\addto\captionsamerican{\renewcommand{\corollaryname}{Corollary}}
\addto\captionsamerican{\renewcommand{\definitionname}{Definition}}
\addto\captionsamerican{\renewcommand{\lemmaname}{Lemma}}
\addto\captionsamerican{\renewcommand{\propositionname}{Proposition}}
\addto\captionsamerican{\renewcommand{\remarkname}{Remark}}
\addto\captionsamerican{\renewcommand{\theoremname}{Theorem}}
\addto\captionsenglish{\renewcommand{\corollaryname}{Corollary}}
\addto\captionsenglish{\renewcommand{\definitionname}{Definition}}
\addto\captionsenglish{\renewcommand{\lemmaname}{Lemma}}
\addto\captionsenglish{\renewcommand{\propositionname}{Proposition}}
\addto\captionsenglish{\renewcommand{\remarkname}{Remark}}
\addto\captionsenglish{\renewcommand{\theoremname}{Theorem}}
\providecommand{\corollaryname}{Corollary}
\providecommand{\definitionname}{Definition}
\providecommand{\lemmaname}{Lemma}
\providecommand{\propositionname}{Proposition}
\providecommand{\remarkname}{Remark}
\providecommand{\theoremname}{Theorem}

\addto\captionsamerican{\renewcommand{\corollaryname}{Corollary}}
\addto\captionsamerican{\renewcommand{\definitionname}{Definition}}
\addto\captionsamerican{\renewcommand{\lemmaname}{Lemma}}
\addto\captionsamerican{\renewcommand{\propositionname}{Proposition}}
\addto\captionsamerican{\renewcommand{\remarkname}{Remark}}
\addto\captionsamerican{\renewcommand{\theoremname}{Theorem}}
\addto\captionsenglish{\renewcommand{\corollaryname}{Corollary}}
\addto\captionsenglish{\renewcommand{\definitionname}{Definition}}
\addto\captionsenglish{\renewcommand{\lemmaname}{Lemma}}
\addto\captionsenglish{\renewcommand{\propositionname}{Proposition}}
\addto\captionsenglish{\renewcommand{\remarkname}{Remark}}
\addto\captionsenglish{\renewcommand{\theoremname}{Theorem}}
\providecommand{\corollaryname}{Corollary}
\providecommand{\definitionname}{Definition}
\providecommand{\lemmaname}{Lemma}
\providecommand{\propositionname}{Proposition}
\providecommand{\remarkname}{Remark}
\providecommand{\theoremname}{Theorem}

\begin{document}

\title{Boundary effects and the stability of the low energy spectrum of the AKLT model}
 \author{Simone Del Vecchio\footnote{Dipartimento di Matematica, Universit\`a degli Studi di Bari, Italy / email: simone.delvecchio@uniba.it}\,,  J\"urg Fr\"ohlich\footnote{Institut f\"ur Theoretiche Physik, ETH-Z\"urich , Switzerland / email: juerg@phys.ethz.ch}\,, Alessandro Pizzo\footnote{Dipartimento di Matematica, Universit\`a di Roma ``Tor Vergata", Italy
/ email: pizzo@mat.uniroma2.it}
, and Alessio Ranallo 
\footnote{Section de math\'ematiques,
Universit\'e de Gen\`eve, Switzerland / email: Alessio.Ranallo@unige.ch}}
  \date{29/10/2024}
\maketitle

\abstract{In this paper we study the low-lying spectrum of the AKLT model perturbed by small, finite-range potentials and with open boundary conditions imposed at the edges of the chain. Our analysis is based  on the \emph{local, iterative Lie Schwinger block-diagonalization method} which allows us to control small interaction terms localized near the boundary of the chain that are
responsible for the possible splitting of the ground-state energy of the AKLT Hamiltonian into energy levels separated
by small gaps. This improves earlier results concerning the persistence of the so called \emph{bulk} gap in these models, 
besides illustrating the power of our general methods in a non-trivial application.}


\section{Introduction}\label{intro}
\setcounter{equation}{0}

In this paper we study finite-range perturbations of the quantum chain known as the AKLT model, 
which was introduced and studied in \cite{AKLT}. Our results concern the low-energy spectrum of the perturbed
models with so called open boundary conditions imposed at the edges of the chain, as studied in \cite{MN}.
 
The main purpose of our work is to devise a general method allowing us to control effects of small 
interaction terms localized near the boundary of the chain that entail the splitting of the ground-state energy of 
the AKLT Hamiltonian into distinct energy levels separated by small gaps. 
Besides offering a new approach to the study of the low-energy spectrum of Hamiltonians of
perturbed AKLT chains, our results improve earlier ones concerning the persistence of the so 
called \emph{bulk} gap, (i.e., the gap between the cluster of energy levels corresponding to the 
four ground-states and the rest of the spectrum of the Hamiltonian separated from these energy 
levels by a uniformly positive gap).

One of the main purposes of our analysis is to show that the \emph{iterative, local Lie-Schwinger block-diagonalization method} 
introduced in \cite{FP} can be applied to small perturbations of Hamiltonians, such as the one of the  AKLT model, with a \textit{multi-dimensional 
ground-state subspace.} It turns out that, in spite of this complication, such models can be analyzed with the help of a \emph{strictly local} 
block-diagonalization method very similar to the one we developed to study quantum chains with a one-dimensional ground-state subspace 
spanned by a product vector.

The key properties of the models studied in this paper enabling us to apply the methods developed in \cite{FP} are the following ones.

\noindent
i)  The expectations of \emph{bulk} observables in the four ground-state vectors of the AKLT chain essentially 
coincide; see Property (\ref{LTQO}), proven in \cite{AKLT}, and generalized under the name of \emph{LTQO} 
condition in \cite{BHM}.

\noindent
ii) A mechanism, involving so-called \textit{Lieb-Robinson bounds,} allowing us to 
treat unperturbed Hamiltonians that are \textit{not} just sums of on-site terms and yet to use \emph{strictly local} 
conjugations as in \cite{FP}. (In the AKLT model the unperturbed Hamiltonian consists of nearest-neighbor interaction 
terms.)

\noindent
In this paper, detailed information on the low-lying spectrum is obtained from local control of effective interaction potentials 
created in the course of our block-diagonalization procedure, including potentials localized near the boundary of the chain.

Our analysis is motivated by recent studies of spectral properties of Hamiltonians appearing in the characterization 
of \textit{``topological phases'';} see, e.g.,  \cite{BN, MZ, BH, BHM, NSY, O1, O2, O3}. 
Various refinements and extensions of the local Lie-Schwinger block-diagonalization method
 can be found in \cite{DFPR1, DFPR2, DFPR3, DFP, DFPRa}. Concerning earlier results on
small perturbations of the AKLT model it should be mentioned that the first proof of stability of the spectral gap 
for Hamiltonians with periodic boundary conditions can be traced back to work by Yarotsky \cite{Y}, who uses 
a cluster expansion. This result has also been established in \cite{MZ} by using the \emph{spectral flow method}. 
In a paper  by Moon and Nachtergaele \cite{MN}, the persistence of the \emph{bulk} gap is established for open 
boundary conditions by adapting the \emph{spectral flow method} originally devised for  periodic boundary conditions. 
In more recent papers (see \cite{NSY1}, \cite{NSY2}), similar results have been proven for a fairly large class of models 
of infinite spin chains.

\noindent
Concerning the AKLT model in higher dimensions, we mention that, in \cite{LSW}, the Hamiltonian on the hexagonal lattice 
has been proven to be gapped. This result has been extended to so called ``decorated" lattices (see \cite{AYLLN}, \cite{PW1}, \cite{PW2}). 
Stability of the spectral gap against small perturbations  has been proven in \cite{LMY} for a class of decorated AKLT models 
on the hexagonal lattice. We expect that the techniques developed in \cite{DFPR3}
can be adapted to treat such models.

\subsection{Definition of the Model}\label{model-cont}

To introduce some notation used throughout our paper we recapitulate the definition of the AKLT model and 
recall its main features. 

\subsubsection{Definition and properties of the AKLT model}\label{prop-AKLT}
Consider a  one-dimensional lattice $\Lambda \subset \mathbb{Z}$ consisting of $N$ sites. By
${\bf{S}}_i=\left(S_i^1, S_i^2, S_i^3\right)$ we denote the components of the spin-1 spin operators at the site 
$i \in \left\{1,\dots, N\right\}$. The Hilbert space of the AKLT chain is given by 
\begin{equation}\label{tensorprod}
\mathcal{H}_{\Lambda} \equiv \mathcal{H}^{(N)}:= \bigotimes_{j=1}^{N} \mathcal{H}_{j}\,,
\end{equation}
where, for each $j\in \Lambda$, $\mathcal{H}_j\simeq \mathbb{C}^3$ is the Hilbert space of the spin-1 
(three-dimensional) representation of $\mathrm{SU}(2)$. 
A ``local observable'' $A$ is a self-adjoint operator on $\mathcal{H}^{(N)}$ localized in an interval 
$\mathcal{I}\subset \left\{1,\dots, N\right\}$ (an interval is a subset of $\Lambda$ consisting of successive sites), 
meaning that
\begin{equation} \label{suppt}
A \,\text{  acts as the identity on  }\,\, \bigotimes_{j\notin\mathcal{I} }\,\, \mathcal{H}_{j}\,.
\end{equation}
The interval $\mathcal{I}$ appearing in (\ref{suppt}) is denoted by $\text{supp}(A)$ and called the 
``support'' of $A$. 
The Hamiltonian of the AKLT model is defined by 
\begin{equation}\label{AKLT-ham}
H^0_{\Lambda}:=\frac{1}{2} \sum\limits_{i =1}^{N-1} [{\bf{S}}_i\cdot {\bf{S}}_{i+1} + \frac{1}{3} ({\bf{S}}_i\cdot {\bf{S}}_{i+1})^2+\frac{2}{3}].
\end{equation}
This Hamiltonian can be written as\, $H^0_{\Lambda}= \sum\limits_{i =1}^{N-1} H_{i,i+1}$, where $H_{i,i+1}:=\mathcal{P}^{(2)}_{i, i+1}$ and
\begin{equation} \label{Hi}
\mathcal{P}^{(2)}_{i,i+1}:=\frac{{\bf{S}}_{i}\cdot {\bf{S}}_{i+1}}{2}+ \frac{({\bf{S}}_{i}\cdot {\bf{S}}_{i+1})^2}{6} + \frac{1}{3}\,.
\end{equation}
The operator $\mathcal{P}^{(2)}_{i,i+1}$ is the orthogonal projection onto the subspace of 
$\mathcal{H}_i\otimes \mathcal{H}_{i+1}$ carrying the spin-$2$ representation of $\mathrm{SU}(2)$
contained in the tensor product of the spin-1 representations with generarors ${\bf{S}}_{i}$ and ${\bf{S}}_{i+1}$.

Next, we recall various important properties of the AKLT model that 
will be used in the sequel; (see \cite{AKLT} for details and proofs).
\begin{itemize}
\item[i)] For the model with open boundary conditions, the ground-state subspace has dimension $4$, 
independently of the length of the chain. An explicit basis for the ground-state subspace is constructed in 
\cite[Eq. 2.7]{AKLT}. 
\item[ii)]
$H^0_{\Lambda}$ is \textit{frustration free}, i.e., $\{0\}\neq \text{Ker}(H^0_{\Lambda})\subseteq \text{Ker}(H_{i,i+1}), \ \forall i \in \left\{1,\dots, N-1\right\}$. 
\item[iii)]
Let $\mathcal{I}\subset \Lambda= \left\{1,\dots, N\right\}$ be an \emph{interval}. We define a Hamiltonian $H^0_{\mathcal{I}}$ associated with $\mathcal{I}$ by
 \begin{equation}\label{H0loc}
H^0_{\mathcal{I}}:= \sum\limits_{i\in \left\{1,\dots, N-1\right\}\,:\, i, i+1 \in \mathcal{I}} H_{i,i+1}\,,
\end{equation}
and denote by $P^{(-)}_\mathcal{I}$ the projection onto $\text{Ker}(H^0_\mathcal{I})$, which is a 
$4$-dimensional subspace of $\mathcal{H}_{\mathcal{I}}:=\bigotimes_{i \in \mathcal{I}}\mathcal{H}_i$; see point i), above. We set $P^{(+)}_{\mathcal{I}}:= 1 - P^{(-)}_\mathcal{I}$, and we denote by $\text{tr}_\mathcal{I}(\cdot)$ the normalized trace with respect to the ground-state subspace of $H^0_\mathcal{I}$;  if $\mathcal{J}\supseteq \mathcal{I}$ and $A$ is localized in $\mathcal{I}$ then
 \begin{equation}\label{omega}
\text{tr}_\mathcal{J}(P^{(-)}_\mathcal{J}A)= \text{tr}_\mathcal{I}(P^{(-)}_\mathcal{I}A).
\end{equation}
Consequently, (\ref{omega}) allows us to define 
the state $\omega(\cdot)$ by
\begin{equation}\label{def-omega}
\omega(A):=\text{tr}_\mathcal{I}(P_{\mathcal{I}}^{(-)} A),
\end{equation}
for all operators $A$ with $\text{supp}(A) \subset \mathcal{I}$.
\item[iv)]
For all pairs of intervals $\mathcal{J}, \mathcal{I}$, with $\mathcal{J}\supseteq \mathcal{I}$, the following estimate holds 
 \begin{equation}\label{LTQO}
\| P^{(-)}_\mathcal{J}(A-\omega(A))P^{(-)}_\mathcal{J}\| \leq 3^{-d( \mathcal{J}^c,\mathcal{I})+1} \|A\|, \ \forall A \, \text{supported in} \, \mathcal{I}, \, \mathcal{J}\supseteq \mathcal{I},
\end{equation}
where $d( \mathcal{J}^c,\mathcal{I})$ is the distance of the complement of the set  $\mathcal{J}$ in $\Lambda$ from the set $\mathcal{I}$.
\end{itemize}

\begin{rem}
A property analogous to iv) is considered in \cite{MN} for a general class of models  and referred to as \textit{``local topological quantum order''} (\emph{LTQO}) condition (see \cite{BHM}).
\end{rem}

 One of the results established in \cite{AKLT} on the model described above is that the spectral gap above the ground-state energy, 
$\text{inf spec}(H^{0}_{\Lambda}) = 0$, is strictly positive, \textit{uniformly} in the length of the chain.

\begin{thm}\label{AKTL-gap}[see Theorem 2.1 \cite{AKLT}]
There exists an $\varepsilon>0$, independent of the length, $N$, of the chain such that \begin{equation}
(\psi, H^0_{\Lambda}\psi)\geq \varepsilon \|\psi\|^2\,,
\end{equation}
for all $\psi$ belonging to  $\text{Ker}(H^0_{\Lambda})^{\perp}$.
\end{thm}
An important ingredient of the mechanism used to analyze this model (alluded to in point 2, at the beginning of Section \ref{intro}) is a  \textit{Lieb-Robison bound}, which we recall next.
\subsubsection{Lieb-Robinson bounds}\label{L-B bounds}
For the AKLT model, a Lieb-Robinson bound (see \cite{LR}) on the propagation speed of ``observables'' in the 
Heisenberg picture has been proven in \cite{NS}. Using the same notation as  in \cite{NS}, we
consider a one-parameter family of functions, $F_a$, defined by 
$$F_a:[0,\infty)\rightarrow (0,\infty)\quad,\quad F_a(r)=e^{-ar}e^{-\sqrt{r}}\frac{1}{(1+r)^3}\,,\quad  a\geq 0\,,$$
which belong to the class of so-called \textit{$\mathcal{F}$-functions} defined in \cite{NS}; namely they 
have the properties
\begin{itemize}
\item $\|F_a\|:=\sum_{i\in\mathbb{Z}^+}F_a(i)<\infty$;
\item there exists a finite constant $C_a>0$ such that, for all $i,j\in\mathbb{Z}$,
$$\sum_{z\in\mathbb{Z}}F_a(|i-z|)F_a(|z-j|)\leq C_a \cdot F_a(|i-j|)\,;$$
\end{itemize}
see Section 6.1 of \cite{MN}. Let $\big\{\exp(i s H^0_{\mathcal{J}})\,\big| \, s \in \mathbb{R}\big\}$ be the 
one-parameter group generated by the Hamiltonian $H^0_{\mathcal{J}}$, $\mathcal{J}\subseteq \Lambda$; 
then Eq.~$(16)$ of \cite{NS} implies that, for two arbitrary observables $A$ and $B$ localized in intervals 
$\mathcal{I}_1, \mathcal{I}_2$, respectively,
\begin{equation}\label{LR-bound}
\| [\exp(i s H^0_{\mathcal{J}})\, A\,\exp(-i s H^0_{\mathcal{J}}) ,B]\| \leq \frac{4\, \|A\|\cdot \|B\|\cdot \|F_0\|}{C_a}\cdot e^{-a\cdot[d(\mathcal{I}_1,\mathcal{I}_2) - \frac{2\, \|\Phi\|_a \cdot C_a\cdot  |s|}{a}]},
\end{equation}
 where $d(\mathcal{I}_1,\mathcal{I}_2)$ is the distance between the sets $\mathcal{I}_1, \mathcal{I}_2$ and
\begin{equation}\label{def-Phi}
\| \Phi\|_a = \frac{\|H^0_{i,i+1}\|}{F_a(1)}\,,
\end{equation}
which, in the context of this paper, is obviously uniformly bounded in $a$. In the sequel (see Section \ref{hooked-proj}) we will set $a=1$.

\subsubsection{Perturbations of the AKLT Hamiltonian} \label{sec-perturbation}
We consider short-range perturbations of $H^0_{\Lambda}$ given by hermitian matrices acting nontrivially on 
Hilbert spaces $\mathcal{H}_{\mathcal{I}}:= \bigotimes_{j\in\mathcal{I} }\,\, \mathcal{H}_{j}$, where 
$\mathcal{I}\subset \Lambda$.  In order to keep our exposition as simple as possible, we consider 
nearest-neighbour interactions denoted by $V_{i,i+1}$, which we assume to be uniformly bounded; i.e.,  
without loss of generality, 
\begin{equation}\label{norm-V}
\|V_{i,i+1}\|\leq 1\,.
\end{equation} 
We define a perturbed Hamiltonian, $K_{\Lambda}(t)$, as the sum of the AKLT Hamiltonian and a perturbation
proportional to a real coupling constant $t$, namely
\begin{equation}\label{bare}
K_{\Lambda}(t):=H_{\Lambda}^0+t\sum_{i=1}^{N-1}\,V_{i,i+1}\,.
\end{equation}
In our proofs we may and will choose $t$ to be non-negative.

\subsection{Main Result} \label{main}
Our main result is the following theorem proven in Section \ref{final-section} (see Theorem \ref{final-thm}).\\
\noindent
{\bf{Theorem.}}
\textit{
There exists some constant $\bar{t} > 0$ independent of the number $N$ of sites in $\Lambda$ such that, 
for any real coupling constant $t$ with $\vert t \vert < \bar{t}$ and for all $1< N< \infty$,
\begin{enumerate}
\item[(i)]{  the spectrum of $K_{\Lambda}(t)$ is contained  in two disjoint, $t$-dependent regions $\sigma^{+}$ and $\sigma^{-}$ separated 
by a gap $\Delta_{\Lambda}(t) \geq \frac{\varepsilon}{4}$, with $\varepsilon$ independent of $N$, as specified in Theorem 1.2; 
i.e., $E'-E''>\Delta_{\Lambda}(t)$, for all $E'\in \sigma^{+}$ and all $E''\in \sigma^{-} $;}
\item[(ii)] for any $d\in \mathbb{N} \cap [1\,,\,\frac{N}{2})$, the eigenspace corresponding to the eigenvalues contained
 in $\sigma^{-}$ is four-dimensional; the gaps between the eigenvalues in $\sigma^{-}$ coincide with the gaps between the 
 eigenvalues of the symmetric matrix
\begin{equation}\label{formula second result}
P^{(-)}_{\Lambda}\,\Big(\,t\sum_{i=1}^{d}\,V_{i,i+1}+\,t\,\sum_{i=N-d}^{N-1}V_{i,i+1}\Big)\,P^{(-)}_{\Lambda}\,,
\end{equation} 
up to corrections bounded by $$|t|\cdot 3^{-(d-1)}\,+\,o(|t|)\,.$$
\end{enumerate}
}

\begin{rem}
One can see by considering a simple example that the small gaps due to interactions localized near the boundaries are typically 
of order $\mathcal{O}(|t|)$. As an interaction we take one component (e.g., the $z$-component) of the spin operator at the first site, multiplied 
by $t$. The eigenvalues of the matrix $P^{(-)}_{\Lambda}\,t\,S^z_1\,P^{(-)}_{\Lambda}$ splits into two groups of two eigenvalues each,  
separated by a gap given by $\approx \frac{4\cdot |t|}{3}$, up to corrections exponentially small in the length of the chain.
\end{rem}

\begin{rem}\label{rem-1.8}
We wish to highlight the effectiveness for explicit computations of the mathematically rigorous formula in result (ii) 
above, which reduces the problem of estimating the eigenvalue splitting to a leading-order calculation in (formal) Rayleigh-Schr\"odinger 
perturbation theory, i.e.,  to calculating \emph{matrix elements of bare potentials in the four-dimensional ground state subspace of the 
AKLT Hamiltonian}.  Using the 
``indistinguishability of the ground-state vectors" (see  (\ref{LTQO})) we can neglect all the bare interaction terms located 
\emph{sufficiently far} from the end-points of the chain. This implies that, by  keeping only the $d$ interactions closest to the 
left- and right end-points, respectively, an error in the values of (small) gaps in the energy spectrum of the perturbed AKLT 
Hamiltonian results that is bounded above by $|t|\cdot 3^{-(d-1)}$.  Notice that, already for $d=10$, the factor 
$3^{-(d-1)}\, \approx\, 5 \cdot 10^{-5}$ is tiny, and  only $20$ potentials have to be kept in the sums shown above. 
\end{rem}



\noindent
{\bf{Notation}}

\vspace{0.2cm}
\noindent
1) The symbol ``$\subset$" denotes a strict inclusion of sets; otherwise the symbol  ``$\subseteq$" is used. 
\\

\noindent
2) 
The symbol $\mathcal{I}\cup \{i\}$ indicates a union of sets of sites of the microscopic lattice.
 \\

{\bf{Acknowledgements.}}
A.P. acknowledges support through the MIUR Excellence Department Project awarded to the Department of Mathematics, University of Rome Tor Vergata, CUP E83C18000100006, and also support through GNFM - INDAM.

\setcounter{equation}{0}

\section{The Block-Diagonalization Algorithm}
In this section we describe some important elements of our method of proving the main result, namely an \textit{iterative 
local block-diagonalization} of the Hamiltonians $K_{\Lambda}(t)$. This method has been developed in several 
previous papers referred to below, starting with \cite{FP}. The original method devised in that paper
cannot be applied directly to the Hamiltonians studied in the present paper, for the following reasons: 
\begin{itemize}
\item[i)]  The unperturbed Hamiltonian $H^0_{\Lambda}$  (see (\ref{AKLT-ham})) is \textit{not ultralocal}; rather, 
it is a collection of nearest-neighbour interaction terms $\mathcal{P}^{(2)}_{i,i+1}$ (see (\ref{Hi})). 
\item[ii)] The ground-state subspace of a Hamiltonian $H^0_{\mathcal{I}}$ associated with an arbitrary 
interval $\mathcal{I} \subset \Lambda$ is not one-dimensional; indeed, it is of dimension four, with a basis of 
\textit{matrix product states} described in \cite{AKLT} and enjoying the properties listed in items i)-iv) of 
Section \ref{prop-AKLT}, above.
\end{itemize}
\subsection{Coarse graining}
\noindent
In order to construct the key ingredients of our analysis, namely \textit{local Lie-Schwinger conjugations} serving 
to block-diagonalize the Hamiltonians $K_{\Lambda}(t)$, the perturbation must be split into terms localized in 
$N$-independent intervals.  
Without loss of generality, we assume that $t>0$ and that
\begin{equation}
(N-1)\, \sqrt{t}\quad ,\quad  \sqrt{t^{-1}}\quad \in \mathbb{N} \,.\label{conditions}
\end{equation}
The perturbation will be split into terms $V_{\mathcal{I}_{ 1,  J}}$ 
supported in intervals $\mathcal{I}_{ 1,  J}$ containing a number of sites approximatively equal to $\sqrt{t^{-1}}$ 
 and belonging to a family $\mathfrak{I}$ of intervals introduced in Definition \ref{rects}, below. 

\begin{rem}
We stress that, for all  chains of length smaller than $\sqrt{t^{-1}}$, i.e., $(N-1)\cdot \sqrt{t}<1$, one is able to block-diagonalize 
the Hamiltonian in one shot, using standard perturbation theory (in the form of Lie-Schwinger conjugations), provided 
$t < \overline{t}$, with $\overline{t}>0$ -- iterations are \textit{not} needed. Actually, for a fixed value of $t<\bar{t}$,  the smaller 
the size of the chain the faster is the convergence of the perturbative series. (Thus, the length of the chain does not imply any 
lower bound on the size of the coupling constant $t$, as one might have guessed mistakenly.)
\end{rem}

 It will be convenient to think of a \emph{macroscopic} (finite) lattice with left endpoint $X=1$, right endpoint $X=N$, and 
lattice spacing $\sqrt{t^{-1}}$. The $M^{th}$ site of this lattice is the point
\begin{equation}\label{macro-lattice}
1+(M-1)\sqrt{t^{-1}}\,,\quad\text{with}\quad 1\leq  M\leq (N-1) \sqrt{t}+1\,,
\end{equation}
of the microscopic lattice $\Lambda$.
The set $\mathcal{I}_{K,J}$ is the \emph{interval} (i.e., a subset of $\Lambda$ consisting of successive sites) 
whose endpoints coincide\footnote{In general, if the range, $\kappa$,  of the interaction terms is larger than $1$, 
the  intervals $\mathcal{I}_{K,J}$ are defined in such a way that they overlap, i.e.,  the endpoints coincide with 
 the sites $M=J$ and $M=J+K$ only up to corrections depending on $\kappa$ and, consequently,  
 $K\cdot \sqrt{t^{-1}}$ is its length up to corrections of the order the step length, $1$, of the 
 \emph{microscopic} lattice where the model is defined.} with  the sites $M=J$ and $M=J+K$ of the 
 \emph{macroscopic} lattice. Notice that it can be helpful to think of the sets $\mathcal{I}_{K,J}$ (and of 
 some enlargements of these sets defined later on) as intervals contained in the real line; for examples, 
 see Figures 1, 2, and 3,  which are intended to display the overlap between such sets.  As an interval 
 of the real line, $\mathcal{I}_{K,J}$ has length $K$ in units of $\sqrt{t^{-1}}$.

\begin{defn} \label{rects}
The  elements of the set $\mathfrak{I}$ are the intervals 
$\mathcal{I}_{K,J}$ (see Fig. 1), where 
\begin{equation}
\mathcal{I}_{K,J}:=\{i\in\mathbb{N}: i\in[1,N] \cap [\,1+(J-1)\sqrt{t^{-1}}\,,\,1+(J-1+K)\sqrt{t^{-1}}\,]\}\,,
\end{equation}
with $K,J \in \mathbb{N}$ such that $1+(J-1+K)\sqrt{t^{-1}}\leq N$. Thus the length, $|\mathcal{I}_{K,J}|$, of $\cI_{K,J}$ is
$
K \cdot \sqrt{t^{-1}}
$.
\end{defn}

\begin{rem}
It follows from the above definitions that the set $\mathfrak{I}$ is closed under taking the union of 
two overlapping elements.
\end{rem}
In the following it will be useful to introduce an ordering relation amongst the intervals labeled by the pairs 
$(K,J)$ with the property that shorter intervals precede longer ones. This relation is specified as follows.
\begin{defn}\label{ordering}
The following defines an ordering relation among the pairs $(K,Q)$ labelling the elements of the set $\mathfrak{I}$
(which will be used in this paper):
\begin{equation}
(K,Q)\succ (K',Q')\quad \text{if}\quad  K>K', \quad \text{or},\quad \text{in case}\quad K=K',\quad \text{if}\quad Q>Q'\,.
\end{equation}
The  symbol $(K,Q)_{\mp 1}$ labels the pair preceding/succeeding $(K,Q)$, respectively, in the ordering relation
of Definition \ref{ordering}. For convenience we shall denote the pair preceding $(1,1)$ by $(0,N)$. 
The last pair is $((N-1)\cdot \sqrt{t},1)$.
\end{defn}

The interval $\mathcal{I}_{1, J}$ is the support of the operator
\begin{equation}
\sum_{i\,:\,i,i+1\,\in \mathcal{I}_{1,J}}\,V_{i,i+1}\,.
\end{equation} Thanks to (\ref{norm-V}) and to our definition of the size, $|\mathcal{I}_{1, J}|$,  
of the interval $\mathcal{I}_{1, J}$, namely $|\mathcal{I}_{1, J}|=\sqrt{t^{-1}}$,
the following operator norm estimate  holds,
\begin{equation}\label{est-V-bare}
\|\,\sum_{i\,:\,i,i+1\,\in \mathcal{I}_{1,J}}\,V_{i,i+1}\, \| \leq \sqrt{t^{-1}}\,.
\end{equation}
 After dividing them by $\sqrt{t^{-1}}$, we denote such a collection of potentials by $V_{\mathcal{I}_{1, J}}$; i.e., 
\begin{equation}
V_{\mathcal{I}_{1, J}}:=\frac{1}{\sqrt{t^{-1}}}\,\sum_{i\,:\,i,i+1\,\in \mathcal{I}_{1,J}}\,V_{i,i+1}\,,
\end{equation}
the support of $V_{\mathcal{I}_{1, J}}$ being $\mathcal{I}_{1, J}$. The observation in (\ref{est-V-bare})
implies that $\|\,V_{\mathcal{I}_{1, J}}\,\|\leq 1$.
In order to implement the block-diagonalization procedure, it is convenient to re-write the Hamiltonian $K_{\Lambda}(t)$
using these definitions; i.e.,
\begin{equation}\label{bare-2}
K_{\Lambda}(t)=H_{\Lambda}^0+\sqrt{t}\, \sum_{\mathcal{I}_{1, J}\subset \Lambda}\,V_{\mathcal{I}_{1,J}}\,.
\end{equation} 
The block-diagonalization is based on spectral projections, $P_{\mathcal{I}_{K, Q}}^{(\pm)}$, associated with 
intervals $\mathcal{I}_{K,Q}$, which we define next.
\begin{defn}\label{def-proj}
By $P^{(-)}_{\mathcal{I}}$ we denote the orthogonal projection onto the ground-state subspace of $H^0_{\mathcal{I}}$,  
and we define
\begin{equation}\label{vacuum_i}
  P_{\mathcal{I}}^{(+)} := 1 - P^{(-)}_{\mathcal{I}}\,.
\end{equation}
We will require analogous definitions of projections associated with general subsets of the lattice $\Lambda$.
\end{defn}

\subsection{Recap of the method for ultralocal unperturbed Hamiltonians}
In order to explain how the method in \cite{FP} has to be modified because of specific features of the AKLT model 
(as compared to the models with \emph{ultralocal} Hamiltonians treated in \cite{FP}),  we first observe that  
the procedure presented in that reference is based on an iterative block-diagonalization of the perturbing 
potentials in the Hamiltonian of those models involving Lie-Schwinger conjugations, assuming that the coupling 
constant, $t(=|t|)$, of the perturbation is sufficiently small. In this paper, too, the block-diagonalization 
is implemented with the help of \emph{local} Lie-Schwinger conjugations; and ``local'' means that 
each conjugation involves only operators supported in an interval $\mathcal{I}_{1,J}$ (of length 1 and with 
left endpoint in $J$ in the macroscopic lattice). More precisely, in the notations of Section \ref{model-cont}, the local Hamiltonian 
supported in the interval $\mathcal{I}_{1,J}$ is conjugated by a suitable local unitary operator. Specifically,
\begin{equation}
H^0_{\mathcal{I}_{1,J}}+\sqrt{t}\,V_{\mathcal{I}_{1,J}}\,,
\end{equation}
(where $H^{0}_{\mathcal{I}_{1,J}}$ is defined in (\ref{H0loc})) is conjugated by a suitably defined unitary operator 
$e^{Z_{\mathcal{I}_{1,J}}}$,
\begin{equation}\label{LieS}
e^{Z_{\mathcal{I}_{1,J}}}(H^0_{\mathcal{I}_{1,J}}+\sqrt{t}\,V_{\mathcal{I}_{1,J}})e^{-Z_{\mathcal{I}_{1,J}}}=H^0_{\mathcal{I}_{1,J}}+\sqrt{t}\,V'_{\mathcal{I}_{1,J}}\,,
\end{equation}
with the purpose to render the new potentials $V'_{\mathcal{I}_{1,J}}\equiv V'_{\mathcal{I}_{1,J}}(t)$ block-diagonal 
with respect to the projections $P^{(-)}_{\mathcal{I}_{1,J}}\,,\, P^{(+)}_{\mathcal{I}_{1,J}}$; 
see Definition \ref{def-proj}. 

\noindent
Obviously new effective interaction potentials are created  as a byproduct of the block diagonalization of the potentials 
$V_{\mathcal{I}_{1,J}}$. Such new potentials are supported in intervals given by connected unions of intervals $\mathcal{I}_{1,Q}$. 
Hence,  in general, a sequence of further conjugations of the Hamiltonian $K_{\Lambda}(t)$ must be introduced 
in order to block-diagonalize the effective interactions created in previous steps, which are supported in ever 
larger intervals $\mathcal{I}_{K,Q}$; (see Definition \ref{rects}).  

In the algorithm designed in \cite{FP}, the steps of the block-diagonalization are indexed by pairs $(K,Q)$ 
labelling the intervals $\mathcal{I}_{K,Q}$ (for which we have introduced an ordering relation in Definition 
\ref{ordering});  that is, in step $(K,Q)$, the potential, $V_{\mathcal{I}_{K,Q}}^{\,(K,Q)_{-1}}$ -- which is 
the potential obtained in the previous step (i.e., in step $(K,Q)_{-1}$) and is supported in $\mathcal{I}_{K,Q}$ -- 
gets block-diagonalized. In the process new terms, given by
\begin{equation}\label{new-term}
\sum_{n=1}^{\infty}\frac{1}{n!}\,ad^{n}Z_{\mathcal{I}_{K,Q}}(V_{\mathcal{I}_{K',Q'}}^{\,(K,Q)_{-1}})\,,
\end{equation}
(where $ad$ stands for \textit{adjoint action}; check (\ref{def-AD}) for its definition) are created that contribute to new interaction potentials, $V^{(K,Q)}_{\mathcal{I}_{K',Q'}\cup \mathcal{I}_{K,Q}}$,   
supported in larger intervals $\mathcal{I}_{K',Q'}\cup \mathcal{I}_{K,Q}$. All the terms created by the 
block-diagonalization procedure with support in the interval 
$\mathcal{I}_{K'',Q''}:=\mathcal{I}_{K',Q'}\cup \mathcal{I}_{K,Q}$ are lumped together. 
To control the size of $V_{\mathcal{I}_{K'',Q''}}$ one has to count certain growth processes 
(of intervals) yielding a given interval $\mathcal{I}_{K'',Q''}:=\mathcal{I}_{K',Q'}\cup \mathcal{I}_{K,Q}$.  
The number of such growth processes can easily be estimated to be at most exponential, i.e., to be
bounded above by $C^{K''}$, for some universal 
constant $C>1$. Since
\begin{equation}\label{est-V}
\|(\ref{new-term})\|\leq \mathcal{O}( \sqrt{t} \cdot \| V_{\mathcal{I}_{K,Q}}^{\,(K,Q)_{-1}}\|\cdot \| V_{\mathcal{I}_{K',Q'}}^{\,(K,Q)_{-1}}\|\,)\,,
\end{equation} 
it is then quite easy to inductively prove a bound of the type
\begin{equation}\label{induc-V-est}
\|V_{\mathcal{I}_{K'',Q''}}^{\,(K,Q)}\|\leq  |t|^{\,\rho  \cdot (K''-1)}\,,
\end{equation}
for some constant $\rho$, with $0<\rho<\frac{1}{2}$, provided that $|t|$ is small enough, uniformly in the number
$N$ of sites of the chain. In the following, we will always assume (w.l.o.g.) that $t\geq 0$, and our results
will hold under the assumption that $t <\bar{t}$, for some constant $\bar{t}$ \textit{independent} of $N (=|\Lambda|)$.

\begin{rem}
In this paper, the term \textit{``step''} can have two different meanings; namely
\begin{enumerate}
\item it can be a label of Hamiltonians and potentials defined in the course of the block-diagonalization procedure: $K_\Lambda^{(K,Q)}(t)$ is the Hamiltonian created in \textit{step} $(K,Q)$ of the block-
diagonalization procedure; 
\item it can mean the iteration step from $(K,Q)_{-1}$ to $(K,Q)$, (i.e., from a certain level $(K,Q)_{-1}$ to the next
one, $(K, Q)$) in the block-diagonalization procedure.
\end{enumerate}

\end{rem}

\subsection{Modifications of the procedure for AKLT-type models}\label{modifs}
Before describing the structure of the Hamiltonian obtained in each step of the block-diagonali-zation 
procedure (see the definitions contained in Section \ref{conj-formulae}), we discuss some new ingredients 
incorporated into the algorithm (see Section \ref{transf}) yielding the new potentials in each step 
of the block-diagonalization. The need for new ingredients becomes apparent already in the steps performed
to block-diagonalize the bare potentials $V_{\mathcal{I}_{1,J}}$ .
The conjugation of the Hamiltonian $K_{\Lambda}$ by the unitary operator $e^{Z_{\mathcal{I}_{1,J}}}$, $1< J< N\cdot \sqrt{t}+1$,
has the effect to not only ``hook up''  to bare interaction potentials, for example $\sqrt{t}\,V_{\mathcal{I}_{1,J-1}}$ 
and $\sqrt{t}\,V_{\mathcal{I}_{1,J+1}}$, but to also ``hook up'' to terms of the unperturbed Hamiltonians   
$H^0_{\mathcal{I}_{1,J-1}}\,,\, H^0_{\mathcal{I}_{1,J+1}}$, namely to the two projections
\begin{equation}
\mathcal{P}^{(2)}_{i_{-}-1,i_{-}}\quad,\quad \mathcal{P}^{(2)}_{i_{+},i_{+}+1}\,,
\end{equation}
where $i_{-}$ and $i_{+}$ are the sites of the microscopic lattice corresponding to the endpoints of the interval 
$\mathcal{I}_{1,J}$; hence, in the conjugation, $\mathcal{P}^{(2)}_{i_{-}-1,i_{-}}$ and $\mathcal{P}^{(2)}_{i_{+},i_{+}+1}$  
(that do not belong to the local Hamiltonian $H^0_{\mathcal{I}_{1,J}}$) get ``hooked up" to other terms. 
Indeed, following the strategy  of \cite{FP}, we should define an anti-symmetric matrix $Z_{\mathcal{I}_{1,J}}$ 
in order to block-diagonalize the interaction potential $V_{\mathcal{I}_{1,J}}$ and observe that, in the course
of the conjugation generated by $Z_{\mathcal{I}_{1,J}}$, new terms of the type
\begin{equation}\label{new-term-1}
\sum_{n=1}^{\infty}\frac{1}{n!}\,ad^{n}Z_{\mathcal{I}_{1,J}}(\mathcal{P}^{(2)}_{i,i+1})\,
\end{equation}
are created whenever 
\begin{equation}
\{i\,,\, i+ 1\}\nsubset \mathcal{I}_{1,J}\,\quad \text{and}\quad \{i\,,\, i+ 1\}\cap \mathcal{I}_{1,J}\neq \emptyset\,.
\end{equation}
We refer to this process in the conjugation as a ``hooking'' of $\mathcal{P}^{(2)}_{i,i+1}$ terms.

\emph{We warn the reader that the conjugation used in the block-diagonalization step $(1,J)$
is generated by an anti-symmetric matrix $Z_{\mathcal{I}^*_{1,J}}$ supported in a somewhat larger interval  
$\mathcal{I}_{1,J}^*\supset \mathcal{I}_{1,J}$. In this informal description, we attempt to explain the problems
that would arise in the block-diagonalization if the matrix $Z_{\mathcal{I}_{1,J}}$ supported in the 
interval $\mathcal{I}_{1,J}$ were used.}

The new interaction terms (\ref{new-term-1}) show some important differences as compared to the 
operators in (\ref{new-term}):
\begin{enumerate}
\item Although the support of the new term displayed in (\ref{new-term-1}) coincides with  $\mathcal{I}_{1,J}$, 
up to a single site, the control of its  norm is quite difficult, since the counterpart of (\ref{new-term}) is
\begin{equation}\label{new-term-1-bis}
\sum_{n=1}^{\infty}\frac{1}{n!}\,ad^{n}Z_{\mathcal{I}_{1,J}}(\frac{\mathcal{P}^{(2)}_{i,i+1}}{\sqrt{t}})\,,
\end{equation}
which cannot be estimated, in a manner similar to (\ref{est-V}), in terms of
$$\mathcal{O}( \sqrt{t} \cdot \| V_{\mathcal{I}_{1,J}}^{\,(1,J)_{-1}}\|\cdot \| \frac{\mathcal{P}^{(2)}_{i,i+1}}{\sqrt{t}}\|\,)\,;$$
indeed this might appear to make it impossible to prove an inductive estimate as in (\ref{induc-V-est}).\\
\item Unless a term supported in $\mathcal{I}_{K,Q}\cup \{ i,  i+ 1\}$ is already block-diagonal, it should be lumped to the effective potential $V^{(K,Q)}_{\mathcal{I}_{K',Q'}}$, for some interval $\mathcal{I}_{K',Q'}$ with 
$$\mathcal{I}_{K',Q'}\supset \mathcal{I}_{K,Q}\cup \{ i,  i+ 1\}\,.$$
\end{enumerate}

The complications described here force us to modify the method proposed in \cite{FP}: in the present paper, 
the strict locality of the on-site operators studied in that paper is given up and replaced by a locality property 
expressed in terms of decay properties of the Green functions of the local Hamiltonians $H^0_{\mathcal{I}}$, 
or, equivalently, by \textit{Lieb-Robinson bounds} associated with the one-parameter groups generated by the operators 
$H^0_{\mathcal{I}}$. Locality is exploited in a careful study of the operator in (\ref{new-term-1}), but associated 
with an enlarged interval $\mathcal{I}^*_{1,J}\supset \mathcal{I}_{1,J}$ introduced in 
Definition \ref{setJ*}, below. 
Thus, in order to block-diagonalize an effective potential supported in $\mathcal{I}_{K,Q}$, we shall use a 
\emph{local unperturbed Hamiltonian} with support in a larger interval $\mathcal{I}^*_{K,Q}\supset \mathcal{I}_{K,Q}$.  
The Lieb-Robinson bound in (\ref{LR-bound}) will be used to show that the off-diagonal part of the operator
\begin{equation}\label{off-diagonal}
ad\,Z_{\mathcal{I}^*_{1,J}}(\frac{\mathcal{P}^{(2)}_{i,i+1}}{\sqrt{t}})=[Z_{\mathcal{I}^*_{1,J}}\,,\,\frac{\mathcal{P}^{(2)}_{i,i+1}}{\sqrt{t}}]
\end{equation}
w.r.t. to spectral projections associated with the enlarged interval  $\overline{\mathcal{I}^*_{1,J}}$ (introduced in  Definition \ref{overline-setJ*}),
\begin{equation}
P^{(-)}_{\overline{\mathcal{I}^*_{1,J}}}\quad,\quad P^{(+)}_{\overline{\mathcal{I}^*_{1,J}}}:=1-P^{(-)}_{\overline{\mathcal{I}^*_{1,J}}}\,,
\end{equation}  
has a norm that decays in $t$ at least as fast as
$$\mathcal{O}( \sqrt{t} \cdot \| V_{\mathcal{I}_{1,J}}^{\,(1,J)_{-1}}\|\,)\,.$$ 
Here one uses that the distance between $\mathcal{I}_{K,Q}$ and the endpoints of $\mathcal{I}^*_{K,Q}$
is of order $\sqrt{t^{-1}}$. This enables us to lump this term, as well as the terms corresponding to 
$n\geq 2$ in the expression 
\begin{equation}\label{new-term-1-bis-bis}
\sum_{n=1}^{\infty}\frac{1}{n!}\,ad^{n}Z_{\mathcal{I}^*_{1,J}}(\frac{\mathcal{P}^{(2)}_{i,i+1}}{\sqrt{t}})\,,
\end{equation}
together with a potential term supported in a larger interval containing $\mathcal{I}^*_{1,J}$,
which will be 
block-diagonalized in a subsequent step. As for the diagonal part of the operator in (\ref{off-diagonal}),  
no extra power of $t$ is gained from the argument based on the Lieb-Robinson bounds; but this is not a problem, because this term 
does not need to be block-diagonalized anymore.
\\

Thanks to the property in (\ref{LTQO}), the use of enlarged intervals also solves a problem\footnote{In the 
following we only try to convey the main ideas  underlying our modification of the block-diagonalization procedure.}  related to the 
degeneracy of the ground-state eigenvalue of Hamiltonians of the type $H^0_{\mathcal{I}}$:
the new potential, which we will denote by $V^{(K,Q)}_{\overline{\mathcal{I}^*_{K,Q}}}$, is supported in the 
enlarged interval $\overline{\mathcal{I}^*_{K,Q}}$ after the block-diagonalization of the potential 
$V^{(K,Q)_{-1}}_{\mathcal{I}_{K,Q}}$ and is not given by the full-fledged Lie-Schwinger series 
(see (\ref{L-S})) associated with the conjugation $e^{Z_{\mathcal{I}^*_{K,Q}}}$. To be more explicit, 
the potential $V^{(K,Q)}_{\overline{\mathcal{I}^*_{K,Q}}}$ 
will contain the following contributions: 
\begin{itemize}
\item[i)]
The expression
\begin{equation}
\omega(V^{(K,Q)_{-1}}_{\mathcal{I}_{K,Q}})+P^{(+)}_{\overline{\mathcal{I}^*_{K,Q}}}\,P^{(+)}_{\mathcal{I}^*_{K,Q}}\,\,\Big[V^{(K,Q)_{-1}}_{\mathcal{I}_{K,Q}}-\omega(V^{(K,Q)_{-1}}_{\mathcal{I}_{K,Q}})\Big]\,P^{(+)}_{\mathcal{I}^*_{K,Q}}\,P^{(+)}_{\overline{\mathcal{I}^*_{K,Q}}}\,,
\end{equation}
which originates in the zero-order term in the Lie Schwinger series, i.e., in
\begin{equation}\label{zero-order}
P^{(-)}_{\mathcal{I}^*_{K,Q}}\,\,V^{(K,Q)_{-1}}_{\mathcal{I}_{K,Q}}\,P^{(-)}_{\mathcal{I}^*_{K,Q}}+P^{(+)}_{\mathcal{I}^*_{K,Q}}\,\,V^{(K,Q)_{-1}}_{\mathcal{I}_{K,Q}}\,P^{(+)}_{\mathcal{I}^*_{K,Q}}\,,
\end{equation}
from which we extract
\begin{equation}\label{extract}
\omega(V^{(K,Q)_{-1}}_{\mathcal{I}_{K,Q}})\,\Big(1-P^{(+)}_{\mathcal{I}^*_{K,Q}}\Big)+P^{(+)}_{\mathcal{I}^*_{K,Q}}\,\,V^{(K,Q)_{-1}}_{\mathcal{I}_{K,Q}}\,P^{(+)}_{\mathcal{I}^*_{K,Q}}
\end{equation}
and neglect a remainder whose norm decays exponentially in the distance, $\mathcal{O}(\sqrt{t^{-1}})$, 
between $\mathcal{I}_{K,Q}$ and the endpoints of $\mathcal{I}^*_{K,Q}$; see (\ref{LTQO}).
This remainder term and the higher-order terms in the Lie-Schwinger series are treated as perturbations and 
lumped together with a potential, supported in a larger interval, that will be block-diagonalized in a later step.
\item[ii)]
The diagonal part w.r.t. $P^{(-)}_{\overline{\mathcal{I}^*_{K,Q}}}, P^{(+)}_{\overline{\mathcal{I}^*_{K,Q}}}$ 
proportional to the projections $\mathcal{P}^{(2)}_{i,i+1}$, hence of terms of the type in (\ref{off-diagonal}).
\end{itemize}
\begin{figure}
\centering
\includegraphics[width=5in, height=1.5in]{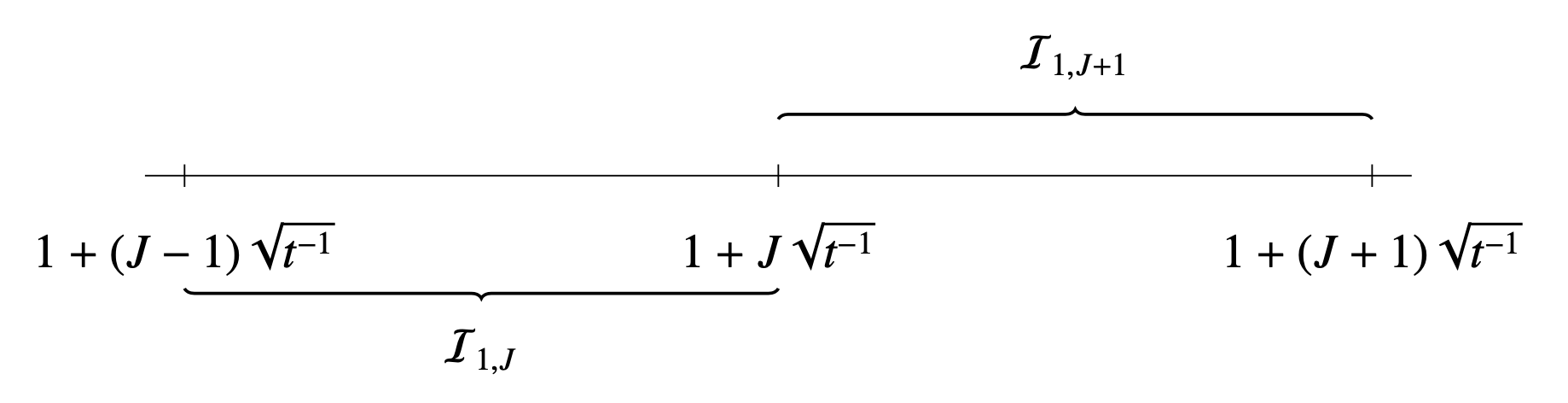}
\caption{The overlapping of the intervals $\mathcal{I}_{1,J}$ and $\mathcal{I}_{1,J+1}$ for $2\leq J \leq (N-1)\sqrt{t}-2$.}
\end{figure}
In replacing (\ref{zero-order}) by (\ref{extract}) we must exclude  from the block-diagonalization steps 
all intervals touching the endpoints of the lattice $\Lambda$. Therefore, we must henceforth
distinguish bulk-  from boundary-potentials, as explained in  
Section \ref{conj-formulae} below. The boundary potentials will get block-diagonalized only 
in the last step that corresponds to the interval $\Lambda$ (given by the entire chain).
\begin{rem}
In all steps of the block-diagonalization except the last one, the local Hamiltonians have a 4-fold degenerate 
ground-state energy. The control of the so-called ``bulk gap'' is however similar to the one used when
considering a Hamiltonian with a non-degenerate ground-state energy. 
\end{rem}

\subsection{Enlarged intervals and unitary conjugations}\label{conj-formulae}

We begin this subsection by introducing enlarged intervals that will be needed in our procedure, 
as explained in Section \ref{modifs}; see also Figures 2 and 3.
\begin{defn}\label{setJ*}
$\mathfrak{I}^*$ is the set of intervals whose elements are the intervals $\mathcal{I}^*_{K,Q}$ defined by
\begin{equation}
\mathcal{I}^*_{K,J}:=\{i\in\mathbb{N}: i\in [1,N] \cap [\,1+(J-\frac{4}{3})\sqrt{t^{-1}},\,1+(J-\frac{2}{3}+K)\sqrt{t^{-1}}\,]\}\,,
\end{equation}
with $K,J$ such that $\cI_{K,J}\in\mathfrak{I}$.
\end{defn} 
\begin{figure}
\centering
\includegraphics[width=5in, height=1.5in]{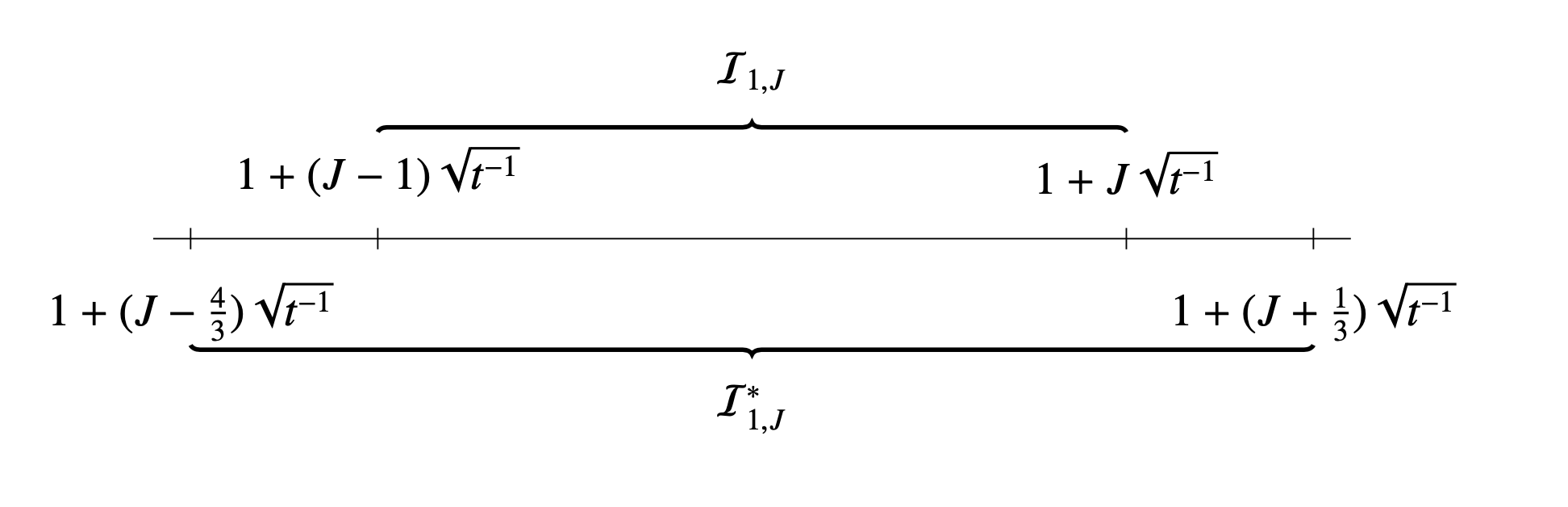}
\caption{The interval $\mathcal{I}_{1,J}$ and its enlargement $\mathcal{I}^\ast_{1,J}$.}
\end{figure}
\begin{defn}\label{enl-tilde}
With each interval $\mathcal{I}^*_{K,Q}\in \mathfrak{I}^*$ we associate an interval $\widetilde{\mathcal{I}}^*_{K,Q}\in \mathfrak{I}$ defined as the smallest interval of type  $\mathcal{I}_{K',Q'}$  containing the interval $\mathcal{I}^*_{K,Q}$.
\end{defn}
\begin{defn}\label{overline-setJ*}
With each interval $\mathcal{I}^*_{K,Q}\in \mathfrak{I}^*$ we associate an interval $\overline{\mathcal{I}^*_{K,Q}}$ 
defined as the interval obtained from $\mathcal{I}^*_{K,Q}$ by including (if present) the two sites in the microscopic lattice, 
nearest to $\mathcal{I}^*_{K,Q}$, one on the right and one on the left.
\end{defn}



In order to implement the block-diagonalization steps,  we define two subsets of the set $\mathfrak{I}$ of
intervals introduced in Definition \ref{rects} of Section \ref{sec-perturbation}:
\begin{eqnarray}
&\mathfrak{I}_{\text{bulk}}:=\{\cI_{K,J}\in\mathfrak{I}: 1,N\notin \cI_{K,J}\}\\
&\mathfrak{I}_{\text{b.ry}}:=\{\cI_{K,J}\in\mathfrak{I}: 1\in \cI_{K,J} \text{ or } N\in \cI_{K,J}\}.
\end{eqnarray}
\begin{defn}[{\textit{Restricted ordering}}]\label{restricted}
The block-diagonalization steps will be associated with  intervals $\cI_{K,Q}\in \mathfrak{I}_{\text{bulk}}$. We will
make use of the ordering introduced in Definition \ref{ordering} (Section \ref{sec-perturbation}) restricted to pairs 
$(K,Q)$ with $\cI_{K,Q}\in \mathfrak{I}_{\text{bulk}}$. Thus the symbols $(K,Q)_{-1}$, $(K,Q)_{+1}$ refer 
to the preceding and the successive element of $(K,Q)$, respectively,  with respect to this restricted ordering, 
i.e., the interval with coordinates $(K,Q)_{-1}$ or $(K,Q)_{+1}$ is required to belong to $\mathfrak{I}_{\text{bulk}}$.
\end{defn}
Using successive unitary conjugations, we shall derive a transformed Hamiltonian that, in step $(K,Q)$, 
will coincide with the operator
\begin{eqnarray}
K_{\Lambda}^{\,(K,Q)}(t)
&:= &H^0_{\Lambda}+\label{fullHamiltonian}\\
& &+{\sqrt{t} }\sum_{Q'}V^{\,(K,Q)}_{\overline{\mathcal{I}^*_{1,Q'}}}+\dots+{\sqrt{t} }\sum_{Q'\,;\, (K,Q')\preceq (K,Q)}V^{\,(K,Q)}_{\overline{\mathcal{I}^*_{K,Q'}}}+\quad\quad\quad \nonumber \\
& &+{\sqrt{t} }\sum_{Q'\,;\, (K,Q')\succ (K,Q)}V^{\,(K,Q)}_{\mathcal{I}_{K,Q'}}+{\sqrt{t} }\sum_{Q'}
V^{\,(K,Q)}_{\mathcal{I}_{K+1,Q'}}+\dots+
{\sqrt{t} }V^{\,(K,Q)}_{\mathcal{I}_{(N-1)\cdot \sqrt{t}-2,2}}\label{K-2}+\\
& &+{\sqrt{t} }\sum_{Q'}W^{\,(K,Q)}_{\mathcal{I}_{1,Q'}}+\dots+
{\sqrt{t} }W^{\,(K,Q)}_{\mathcal{I}_{(N-1)\cdot \sqrt{t},1}}\,,\label{K-3}
\end{eqnarray}
where the two types of potentials, ``V'' and ``W,'' are specified below:
\begin{figure}
\centering
\includegraphics[width=5in, height=1.9in]{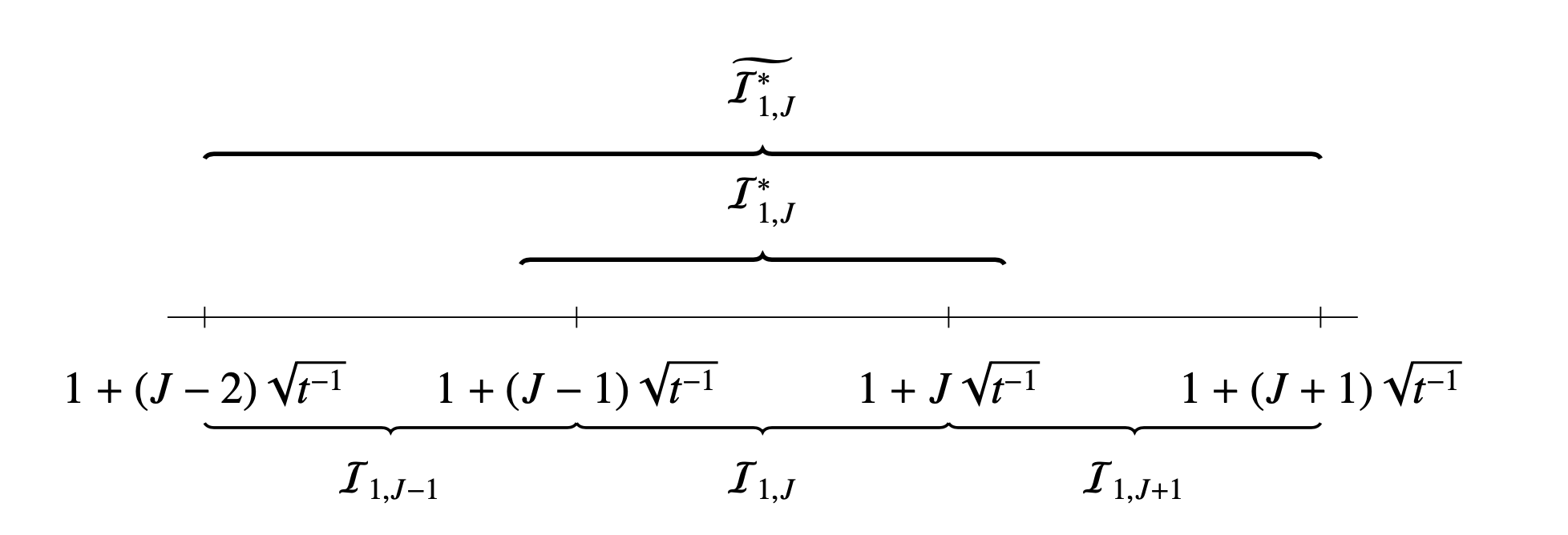}
%
\caption{How an interval $\mathcal{I}_{1,J}$ relates to $\mathcal{I}^\ast_{1,J}$ and $\widetilde{\mathcal{I}}^\ast_{1,J}$.}
\end{figure}

\begin{itemize}
\item
Depending on whether $(K',Q')\preceq (K,Q)$ or $(K',Q')\succ (K,Q)$ potentials of type ``V'' are labeled by intervals 
$\overline{\mathcal{I}^*_{K',Q'}}$ or  by  intervals $\mathcal{I}_{K',Q'}$, respectively; in both cases $\mathcal{I}_{K',Q'}\in \mathfrak{I}_{\text{bulk}}$. The first type of potentials, i.e., those corresponding to $(K',Q')\preceq (K,Q)$, are block-diagonalized, and the block-diagonalization is w.r.t. the two projections $P^{(-)}_{\overline{\mathcal{I}^*_{K',Q'}}}$, $P^{(+)}_{\overline{\mathcal{I}^*_{K',Q'}}}$ (see Definition \ref{def-proj}); more precisely, they are of the form
\begin{equation}
V^{\,(K,Q)}_{\overline{\mathcal{I}^*_{K',Q'}}}=P^{(+)}_{\overline{\mathcal{I}^*_{K',Q'}}}\,
V^{\,(K',Q')}_{\overline{\mathcal{I}^*_{K',Q'}}}\,P^{(+)}_{\overline{\mathcal{I}^*_{K',Q'}}}+P^{(-)}_{\overline{\mathcal{I}^*_{K',Q'}}}\,
V^{\,(K',Q')}_{\overline{\mathcal{I}^*_{K',Q'}}}\,P^{(-)}_{\overline{\mathcal{I}^*_{K',Q'}}}\,.
\end{equation}
It is straightforward to check that they are block-diagonal w.r.t. any pair $P^{(+)}_{\mathcal{I}}\,,\,P^{(-)}_{\mathcal{I}}$ with $\mathcal{I}\supset \mathcal{I}^*_{K',Q'}$, due to the \textit{frustration free} property of $H^0_{\Lambda}$.
\item
The potentials $W^{\,(K,Q)}_{\mathcal{I}_{K',Q'}}$ are characterized by the property that $\mathcal{I}_{K',Q'}\in \mathfrak{I}_{\text{b.ry}}$, i.e., they are zero if $\mathcal{I}_{K',Q'}\notin \mathfrak{I}_{\text{b.ry}}$. They get block-diagonalized only in the very last step.
\end{itemize}
\noindent
The way these potentials are produced in each step of the block-diagonalization procedure is explained in Sect. \ref{transf}. 
The procedure has the property that, in step $ (K,Q)$, the potential $V^{\,(K,Q)_{-1}}_{\mathcal{I}_{K,Q}}$ 
is transformed to a potential $V^{\,(K,Q)}_{\overline{\mathcal{I}^*_{K,Q}}}$ \emph{related} to  the Lie-Schwinger series
(for details see point b) in Definition \ref{def-interactions-multi}):
\begin{equation}\label{L-S}
\sum_{j=1}^{\infty}t^{\frac{j-1}{2}}\,(V^{(K,Q)_{-1}}_{\mathcal{I}^*_{K,Q}})^{\text{diag}}_j\,.
\end{equation}
The operators $(V^{(K,Q)_{-1}}_{\mathcal{I}^*_{K,Q}})_j^{\text{diag}}$ will be defined below, and ``$\text{diag}$'' stands for the diagonal part w.r.t. to the two projections $P^{(-)}_{\mathcal{I}^*_{K,Q}}\,,\,P^{(-)}_{\mathcal{I}^*_{K,Q}}$; they are
determined by
\begin{equation}
e^{Z_{\mathcal{I}^*_{K,Q}}}\,(G_{\mathcal{I}^*_{K,Q}}+{\sqrt{t} }V_{\mathcal{I}_{K,Q}}^{\,(K,Q)_{-1}})\,e^{-Z_{\mathcal{I}^*_{K,Q}}}=:G_{\mathcal{I}^*_{K,Q}}+{\sqrt{t} }\sum_{j=1}^{\infty}t^{\frac{j-1}{2}}\,(V^{(K,Q)_{-1}}_{\mathcal{I}^*_{K,Q}})^{\text{diag}}_j\,,
\end{equation}
where
 \begin{eqnarray}\label{expression-G}
G_{\mathcal{I}^*_{K,Q}}&:=& H_{\mathcal{I}^*_{K,Q}}^{0}+{\sqrt{t} }\sum_{J=1}^{K-1}\,\,\sum_{\overline{\mathcal{I}^*_{J,Q'}}\subset \mathcal{I}^*_{K,Q}} V^{(K,Q)_{-1}}_{\overline{\mathcal{I}^*_{J,Q'}}}\,\,.
\end{eqnarray} 
\noindent
The reader should notice that the (second) sum on the right side of (\ref{expression-G}) does \textit{not} include those intervals $\mathcal{I}^\ast_{J,Q'}$ that share one of their endpoints with $\mathcal{I}^\ast_{K,Q}$. As a consequence, $G_{\mathcal{I}^*_{K,Q}}$ is localized in $\mathcal{I}^*_{K,Q}$.
The operator $Z_{\mathcal{I}^*_{K,Q}}$ is given by
\begin{equation}\label{S-definition-1}
Z_{\mathcal{I}^*_{K,Q}}:=\sum_{j=1}^{\infty}{t^{\frac{j}{2}}}(Z_{\mathcal{I}^*_{K,Q}})_j\,
\end{equation}
where the terms $(Z_{\mathcal{I}^*_{K,Q}})_j$ are defined recursively as follows: 
\begin{itemize}
\item
\begin{equation}
\label{S-definition-2-0}
(Z_{\mathcal{I}^*_{K,Q}})_j
:=\frac{1}{G_{\mathcal{I}^*_{K,Q}}-E_{\mathcal{I}^*_{K,Q}}}P^{(+)}_{\mathcal{I}^*_{K,Q}}\,(V^{(K, Q)_{-1}}_{\mathcal{I}^*_{K,Q}})_j\,P^{(-)}_{\mathcal{I}^*_{K,Q}}-h.c.\,,
\end{equation}
where 
\begin{equation}\label{def-E}
E_{\mathcal{I}^*_{K,Q}}:=\,{\sqrt{t} }\sum_{J=1}^{K-1}\,\,\sum_{\overline{\mathcal{I}^*_{J,Q'}}\subset \mathcal{I}^*_{K,Q}} \,\omega( V^{(K,Q)_{-1}}_{\overline{\mathcal{I}^*_{J,Q'}}})
\end{equation}
and $\omega$ is defined in (\ref{def-omega});
\item
\begin{equation}\label{formula-V_1}
(V^{(K,Q)_{-1}}_{\mathcal{I}^*_{K,Q}})_1:=V^{(K,Q)_{-1}}_{\mathcal{I}_{K,Q}}\,,
\end{equation}
and, for $j\geq 2$,
\begin{eqnarray}
&&(V^{(K,Q)_{-1}}_{\mathcal{I}^*_{K,Q}})_j:=\nonumber\\
&&\sum_{p\geq 2, r_1\geq 1 \dots, r_p\geq 1\, ; \, r_1+\dots+r_p=j}\frac{1}{p!} ad\,(Z_{\mathcal{I}^*_{K,Q}})_{r_1}\Big(ad\,(Z_{\mathcal{I}^*_{K,Q}})_{r_2}\dots (ad\,(Z_{\mathcal{I}^*_{K,Q}})_{r_p}(G_{\mathcal{I}^*_{K,Q}}))\dots\Big)+\quad \nonumber \\
&&\sum_{p\geq 1, r_1\geq 1 \dots, r_p\geq 1\, ; \, r_1+\dots+r_p=j-1}\frac{1}{p!}
ad\,(Z_{\mathcal{I}^*_{K,Q}})_{r_1}\Big(ad\,(Z_{\mathcal{I}^*_{K,Q}})_{r_2}\dots (ad\,(Z_{\mathcal{I}^*_{K,Q}})_{r_p}
((V^{(K,Q)_{-1}}_{\mathcal{I}^*_{K,Q}})_1))\dots \Big)\,,\quad\quad\quad \label{formula-V_j}
\end{eqnarray}
where the adjoint action of an operator $A$ on an operator $B$ is defined by
\begin{equation}\label{def-AD}
ad\, A\,(B):=\big[A\,,\,B\big]\,,
\end{equation}
and, recursively,
\begin{equation}
{ad}^n A\,(B):=\big[A\,,\,{ad}^{n-1} A\,(B)\big]\,, \text{  for  }\, n\geq 2\,. 
\end{equation}
\end{itemize}
We note that the construction of $Z_{\mathcal{I}^*_{K,Q}}$ requires control of the spectral gap of 
$G_{\mathcal{I}^*_{K,Q}}$ above the ground-state energy, i.e., an estimate on
$$\inf \text{spec} \,[(G_{\mathcal{I}^*_{K,Q}}-E_{\mathcal{I}^*_{K,Q}})P^{(+)}_{\mathcal{I}^*_{K,Q}}]\,,$$ 
which we will outline in Section \ref{gap}. 
\begin{rem}
The reader is invited to notice that the operators of type ``W''  are not included in the definition of the Hamiltonian $G_{\mathcal{I}^*_{K,Q}}$. 
\end{rem}
\begin{rem}
The Lie-Schwinger series and, accordingly, the series defining $Z_{\mathcal{I}^*_{K,Q}}$ could actually be
truncated, thanks to the structure of the algorithm specified in Section \ref{transf} below.
\end{rem}




\subsection{The Algorithm}\label{transf}
 
The following definitions iteratively specify two families of effective interaction potentials, 
$V^{(K,Q)}_{\cI_{R,J}}$ and $W^{(K,Q)}_{\cI_{R,J}}$, and we note that the step of the algorithm 
labelled by $(K,Q)$ is such that $\cI_{K,Q}\in \mathfrak{I}_{\text{bulk}}$ and $\cI_{R,J}$ belongs to $\mathfrak{I}_{\text{bulk}}$ if it is the support of $V^{(K,Q)}_{\cI_{R,J}}$ or to  $\mathfrak{I}_{\text{b.ry}}$ if it is the support of $W^{(K,Q)}_{\cI_{R,J}}$.

\begin{defn} \label{pot(0,N)}

\noindent
\begin{itemize}
\item For $\cI_{1,J}\in \mathfrak{I}_{\text{bulk}}$, we define
\begin{equation}
V_{\cI_{1,J}}^{(0,N)}:=\frac{1}{\sqrt{t^{-1}}}\sum_{\{i,i+1\}\subset \cI_{1,J}}V_{i,i+1}\,.
\end{equation}
\item For $\cI_{1,J}\in \mathfrak{I}_{\text{b.dry}}$, we define
\begin{equation}
W_{\cI_{1,J}}^{(0,N)}:=\frac{1}{\sqrt{t^{-1}}}\sum_{\{i,i+1\}\subset \cI_{1,J}}V_{i,i+1}.
\end{equation}
\end{itemize}
Furthermore,
\begin{itemize}
\item for $\cI_{K,J}\in \mathfrak{I}_{\text{bulk}}$, with $K\geq 2$, we define
\begin{equation}
V_{\cI_{K,J}}^{(0,N)}:=0\,,
\end{equation}
\item for $\cI_{K,J}\in \mathfrak{I}_{\text{b.dry}}$, with $K\geq 2$, we define
\begin{equation}
W_{\cI_{K,J}}^{(0,N)}:=0\,.
\end{equation}
\end{itemize}
We view $(0,N)$ as the predecessor of $(1,2)$, in accordance with the restricted ordering 
introduced in Definition \ref{restricted}.
\end{defn} 
\noindent
\textbf{Notation:} In the following, $\omega$ is the state introduced in (\ref{def-omega}); moreover, $i^\ast_{-}$ and
$i^\ast_{+}$ are the two boundary sites in the microscopic lattice of the interval $\mathcal{I}_{K,Q}^\ast$.

\begin{defn}\label{def-interactions-multi}
Assuming that, for an arbitrary $(K,Q)_{-1}$ with $(K,Q)_{-1} \succ (0,N)$, the operators 
$V^{(K,Q)_{-1}}_{\mathcal{I}_{R,J}}\,,\, V^{(K,Q)_{-1}}_{\overline{\mathcal{I}^*_{R,J}}}\,,\,
{W^{(K,Q)_{-1}}_{\mathcal{I}_{R,J}}\,} $ are well defined for any $(R,J)$ in $\mathfrak{I}_{\text{bulk}}$ and in $\mathfrak{I}_{\text{b.dry}}$, respectively, and that the operators
$Z_{\mathcal{I}^*_{K,Q}}$ (see (\ref{S-definition-1}))  are well defined, and assuming that if $(K,Q)=(1,2)$ then 
$Z_{\mathcal{I}^*_{1,2}}$ is well defined, then definitions  a-1), a-2), b), c-1), and c-2) (see below) are meaningful.
\noindent
Such prescriptions are organized into three groups, $\mathcal{A}$, $\mathcal{B}$, and $\mathcal{C}$; for each of them we give first a description in words.
\begin{enumerate}
\item[$\mathcal{A})$] Items a-1) and a-2) below deal with identity maps, that is they describe situations where for a given interval $\mathcal{I}$ the corresponding potential (supported in $\mathcal{I}$) does not change from step $(K,Q)_{-1}$ to step $(K,Q)$; in terms of the conjugation associated with step $(K,Q)$, the new potential is either the zero order term in the expansion (in $Z_{\mathcal{I}^*_{K,Q}}$) of
\begin{equation}
e^{Z_{\mathcal{I}^*_{K,Q}}}\,O_{\mathcal{I}}\,e^{-Z_{\mathcal{I}^*_{K,Q}}}\,,
\end{equation}
where $O_{\mathcal{I}}$ stands for the potential under consideration  in step $(K,Q)_{-1}$, or just the operator $O_{\mathcal{I}}$ whenever $[O_{\mathcal{I}}\,,\,e^{-Z_{\mathcal{I}^*_{K,Q}}}]=0$;
\begin{itemize}
\item[a-1)]
if  $(K,Q)\prec (R,J)$, \,$\cI_{R,J}\in\mathfrak{I}_{\text{bulk}}$ and $\mathcal{I}^*_{K,Q} \nsubseteq \mathcal{I}_{R,J}$
we set
\begin{equation}\label{a-0}
V^{(K,Q)}_{\mathcal{I}_{R,J}}:=V^{(K,Q)_{-1}}_{\mathcal{I}_{R,J}}\,;
\end{equation}
if  $(K,Q)\prec (R,J)$,\, $\cI_{R,J}\in{\mathfrak{I}_{\text{b.ry}}}$ and $\mathcal{I}^*_{K,Q} \nsubseteq \mathcal{I}^*_{R,J}$
we set
\begin{equation}\label{a-1}
{W^{(K,Q)}_{\mathcal{I}_{R,J}}:=W^{(K,Q)_{-1}}_{\mathcal{I}_{R,J}}}\,;
\end{equation}

\item[a-2)]
if  $(K,Q)\succ (R,J)$,  for $\cI_{R,J}\in \mathfrak{I}_{\text{bulk}}$, we set
\begin{equation}\label{a-bis}
V^{(K,Q)}_{\overline{\mathcal{I}^*_{R,J}}}:=V^{(K,Q)_{-1}}_{\overline{\mathcal{I}^*_{R,J}}},
\end{equation}
if  $(K,Q)\succ (R,J)$, for $\cI_{R,J}\in \mathfrak{I}_{\text{b.ry}}$, we set
\begin{equation}\label{a-bis-1}
W^{(K,Q)}_{{\mathcal{I}_{R,J}}}:=W^{(K,Q)_{-1}}_{{\mathcal{I}_{R,J}}};
\end{equation}
\end{itemize}
\item[$\mathcal{B})$] Item  b) below describes the process that takes place when the label $(K,Q)$ of the step coincides with the label $(R,J)$ of the potential under consideration. By construction, only  for potentials of type ``V" the labels $(R,J)$ and $(K,Q)$ can coincide. As anticipated in Section \ref{modifs},  the map which defines the new potential in step $(K,Q)$  consists of two operations:

\noindent
- extracting the quantity (\ref{b1main}) from the leading order term of the Lie-Schwinger series defined in (\ref{L-S}) and associated with the potential supported in $\mathcal{I}_{K,Q}\equiv \mathcal{I}_{R,J}$ in step $(K,Q)_{-1}$;

\noindent
- extracting the diagonal part  from the first order of what we refer to as hooking of the (rescaled) projection terms, i.e., 
\begin{equation}
e^{Z_{\mathcal{I}^*_{K,Q}}}\,\frac{\mathcal{P}^{(2)}_{i-1,i}}{\sqrt{t}}\,e^{-Z_{\mathcal{I}^*_{K,Q}}}-\frac{\mathcal{P}^{(2)}_{i-1,i}}{\sqrt{t}}\,,
\end{equation}
 where the support of $\mathcal{P}^{(2)}_{i-1,i}$ overlaps with $\mathcal{I}^*_{K,Q}$ but it is not contained in it;
\begin{itemize}
\item[b)]
if $(K,Q)= (R,J)$ then
\begin{eqnarray}
V^{(K,Q)}_{\overline{\mathcal{I}^*_{R,J}}}&:= &\omega(V^{(K,Q)_{-1}}_{\mathcal{I}_{K,Q}})+P^{(+)}_{\overline{\mathcal{I}^*_{K,Q}}}\,P^{(+)}_{\mathcal{I}^*_{K,Q}}\,\,\Big[V^{(K,Q)_{-1}}_{\mathcal{I}_{K,Q}}-\omega(V^{(K,Q)_{-1}}_{\mathcal{I}_{K,Q}})\Big]\,P^{(+)}_{\mathcal{I}^*_{K,Q}}\,P^{(+)}_{\overline{\mathcal{I}^*_{K,Q}}}\quad\quad\quad \label{b1main}\\
& &+P^{(+)}_{\overline{\mathcal{I}^*_{R,J}}}\,\Big(adZ_{\mathcal{I}^*_{K,Q}\equiv \mathcal{I}^*_{R,J}}(\,\frac{\mathcal{P}^{(2)}_{i^*_{-}-1,i^*_{-}}}{\sqrt{t}})\Big)\,P^{(+)}_{\overline{\mathcal{I}^*_{R,J}}}\label{b-hop1}\\
& &+P^{(+)}_{\overline{\mathcal{I}^*_{R,J}}}\,\Big(adZ_{\mathcal{I}^*_{K,Q}\equiv \mathcal{I}^*_{R,J}}(\,\frac{\mathcal{P}^{(2)}_{i^*_{+}, i^*_{+}+1}}{\sqrt{t}})\Big)\,P^{(+)}_{\overline{\mathcal{I}^*_{R,J}}} \label{b-hop2}\,,
\end{eqnarray}
where $i^*_{-}$ and $i^*_{+}$ are the sites of the microscopic lattice corresponding to the endpoints of the interval 
$\mathcal{I}^*_{K,Q}\equiv \mathcal{I}^*_{R,J}$\,.
\end{itemize}
\item[$\mathcal{C})$]  Items c-1) and c-2) below describe growth processes for the V and the W potentials, respectively. Regarding  the new potential of type V (see c-1)) associated with a  given fixed interval $\mathcal{I}_{R,J}$ with $\mathcal{I}^*_{K,Q}\subset \mathcal{I}_{R,J}$,  it does not involve W operators. It involves $V$ operators, but also includes possible contributions coming from the hooking (in step $(K,Q)$) of rescaled projections, namely \emph{higher order} and \emph{off-diagonal} first order terms.  The growth process prescribed in c-2) for the W potentials includes all the terms which upon the conjugation of the Hamiltonian in step $(K,Q)$ turned out to be supported in the  interval $\mathcal{I}_{R,J}$ supposed to be in $\mathfrak{I}_{\text{b.ry}}$. Concerning the nontrivial structure designed in c-1) and c-2), the reader is referred to the explanations in Remarks \ref{rem-c-1}, \ref{rem-c-2}, and \ref{rem-c-3}.
\begin{itemize}
\item[c-1)]
if $\mathcal{I}^*_{K,Q} \subset \mathcal{I}_{R,J}\in\mathfrak{I}_{\text{bulk}}$ then
\begin{eqnarray}\label{construction-conn}
V^{(K,Q)}_{\mathcal{I}_{R,J}} &:= & e^{Z_{\mathcal{I}^*_{K,Q}}}\,V^{(K,Q)_{-1}}_{\mathcal{I}_{R,J}}\,e^{-Z_{\mathcal{I}^*_{K,Q}}}\label{identity-c}\\
& &+\sum_{\mathcal{I}_{K',Q'}\in [\mathcal{G}^{(K,Q)}_{\mathcal{I}_{R,J}}]_1}\,\sum_{n=1}^{\infty}\frac{1}{n!}\,ad^{n}Z_{\mathcal{I}^*_{K,Q}}(V^{(K,Q)_{-1}}_{\mathcal{I}_{K',Q'}})\label{A-map-1-bis}\\
& &+\sum_{\mathcal{I}^*_{K',Q'}\in [\mathcal{G}^{(K,Q)}_{\mathcal{I}_{R,J}}]_2}\,\sum_{n=1}^{\infty}\frac{1}{n!}\,ad^{n}Z_{\mathcal{I}^*_{K,Q}}(V^{(K,Q)_{-1}}_{\overline{\mathcal{I}^*_{K',Q'}}})\label{Valgo3}\\
& &+\delta_{\widetilde{\mathcal{I}}^*_{K,Q}=\mathcal{I}_{R,J}} \sum_{\mathcal{I}^*_{K',Q'}\in [\mathcal{G}^{(K,Q)}_{\mathcal{I}_{R,J}}]_3}\,\sum_{n=1}^{\infty}\frac{1}{n!}\,ad^{n}Z_{\mathcal{I}^*_{K,Q}}(V^{(K,Q)_{-1}}_{\overline{\mathcal{I}^*_{K',Q'}}})\label{Valgo67}\\
& &{+\delta_{\widetilde{\mathcal{I}}^*_{K,Q}=\mathcal{I}_{R,J}} [P^{(-)}_{\mathcal{I}^*_{K,Q}}V^{(K,Q)_{-1}}_{\mathcal{I}_{K,Q}}P^{(-)}_{\mathcal{I}^*_{K,Q}}-\omega(V^{(K,Q)_{-1}}_{\mathcal{I}_{K,Q}})P^{(-)}_{\mathcal{I}^*_{K,Q}}] } \,\,\quad\quad \label{expdecayerror}\\ 
& &+\delta_{\widetilde{\mathcal{I}}^*_{K,Q}=\mathcal{I}_{R,J}} [\sum_{m=2}^{\infty}{t^{\frac{m-1}{2}}}(V^{(K,Q)_{-1}}_{\mathcal{I}^*_{K,Q}})^{\text{diag}}_m]\label{lshighorder}\\
& &+\delta_{\widetilde{\mathcal{I}}^*_{K,Q}=\mathcal{I}_{R,J}}\Big(\sum_{n=2}^{\infty}\frac{1}{n!}\,ad^{n}Z_{\mathcal{I}^*_{K,Q}}(\,{\frac{\mathcal{P}^{(2)}_{i^\ast_{-}-1,i^\ast_{-}}}{\sqrt{t}}}+{\frac{\mathcal{P}^{(2)}_{i^\ast_{+},i^\ast_{+}+1}}{\sqrt{t}}})\Big) \label{b-21}\\
&&+\delta_{\widetilde{\mathcal{I}}^*_{K,Q}=\mathcal{I}_{R,J}}[P^{(-)}_{\overline{\mathcal{I}^*_{K,Q}}}\Big(adZ_{\mathcal{I}^*_{K,Q}}(\,{\frac{\mathcal{P}^{(2)}_{i^\ast_{-}-1,i^\ast_{-}}}{\sqrt{t}}}+{\frac{\mathcal{P}^{(2)}_{i^\ast_{+},i^\ast_{+}+1}}{\sqrt{t}}})\Big)P^{(+)}_{\overline{\mathcal{I}^*_{K,Q}}}+h.c.]\,,\quad \label{b-212}
\end{eqnarray}
where 
\begin{eqnarray}
[\mathcal{G}^{(K,Q)}_{\mathcal{I}_{R,J}}]_1 &:=&\Big\{ \, \mathcal{I}_{K',Q'} \in\mathfrak{I}_{\text{bulk}}\,\,\vert \,\,(K',Q')\succ (K,Q)\,,\,\mathcal{I}_{K',Q'}\cap \mathcal{I}^*_{K,Q}\neq \emptyset , \nonumber\\
&&\,\mathcal{I}_{K',Q'}\neq \mathcal{I}_{R,J}\,,\,\text{and} \,\, \widetilde{\mathcal{I}}^*_{K,Q}\cup \mathcal{I}_{K',Q'}=\mathcal{I}_{R,J}\,\,\Big\}\label{intervals}
\end{eqnarray}
\begin{eqnarray}
[ \mathcal{G}^{(K,Q)}_{\mathcal{I}_{R,J}}]_2  &:=& 
\Big\{ \, \mathcal{I}^*_{K',Q'} \in\mathfrak{I}_{\text{bulk}}\,\,\vert \,\,(K,Q)\succ (K',Q')\,,\,\mathcal{I}^*_{K',Q'}\cap \mathcal{I}^*_{K,Q}\neq \emptyset \,, \mathcal{I}^*_{K',Q'}\nsubset\mathcal{I}^*_{K,Q} \nonumber\\ 
&& \text{and} \,\,
\widetilde{\mathcal{I}}^*_{K,Q}\cup \widetilde{\mathcal{I}}^*_{K',Q'}=\mathcal{I}_{R,J}\,\,\Big\} \,\nonumber
\end{eqnarray}
\begin{eqnarray}
[ \mathcal{G}^{(K,Q)}_{\mathcal{I}_{R,J}}]_3 &:=& 
\Big\{ \, \mathcal{I}^*_{K',Q'} \in\mathfrak{I}_{\text{bulk}}\,\,\vert \,\, \mathcal{I}^*_{K',Q'}\subset\mathcal{I}^*_{K,Q},\,  i_-^* \in\mathcal{I}^*_{K',Q'}\text{ or } i_+^*\in\mathcal{I}^*_{K',Q'}\,\Big\} \,\nonumber
\end{eqnarray}
\begin{rem}\label{rem-c-1}
The terms in (\ref{Valgo67}), (\ref{expdecayerror}), and (\ref{lshighorder}) are related to the block-diagonalization in step $(K,Q)$ and are present only if $\widetilde{\mathcal{I}}^*_{K,Q}=\mathcal{I}_{R,J}$. More precisely, we observe that: (\ref{Valgo67}) originates from the definition of (\ref{expression-G}) in the sense that accounts for the higher order terms of the conjugation of  those potentials supported in intervals $\mathcal{I}^\ast_{J,Q'}$, with $\mathcal{I}^*_{J,Q'}\subset \mathcal{I}^*_{K,Q}$,  that share one of their endpoints with $\mathcal{I}^\ast_{K,Q}$; (\ref{expdecayerror}) collects what is left of the first order term of the Lie Schwinger series after extracting the quantity in (\ref{b1main}) which enters the definition in b); (\ref{lshighorder}) is the Lie Schwinger series above first order.
\end{rem}
\begin{rem}\label{rem-c-2}
The companion off-diagonal terms of (\ref{b-hop1}) and (\ref{b-hop2}) respectively are two terms of the type in (\ref{b-212}) for an interval $\mathcal{I}_{R',J',}$  such that  $\mathcal{I}_{R',J'}=\widetilde{\mathcal{I}}^*_{K,Q}=\widetilde{\mathcal{I}}^*_{R,J}$. The term in (\ref{b-21}) accounts for the higher order terms of the operator resulting from the hooking of the projections in step $(K,Q)$, provided $\widetilde{\mathcal{I}}^*_{K,Q}=\mathcal{I}_{R,J}$.
\end{rem}

\item[c-2)]
if $\mathcal{I}^*_{K,Q}\subset \mathcal{I}_{R,J}\in\mathfrak{I}_{\text{b.ry}}$,
\begin{eqnarray}\label{diag-W-growth}
W^{(K,Q)}_{{\mathcal{I}_{R,J}}}&:= &e^{Z_{\mathcal{I}^*_{K,Q}}}\,W^{(K,Q)_{-1}}_{{\mathcal{I}_{R,J}}}\,e^{-Z_{\mathcal{I}^*_{K,Q}}}\label{c-2-off-0}\\
& &+\sum_{\mathcal{I}_{K',Q'}\in \mathcal{G}^{(K,Q)}_{\mathcal{I}_{R,J}}}\,\sum_{n=1}^{\infty}\frac{1}{n!}\,ad^{n}Z_{\mathcal{I}^*_{K,Q}}(W^{(K,Q)_{-1}}_{{\mathcal{I}_{K',Q'}}})\label{c-2-off-2}\\
& &+\sum_{\mathcal{I}_{K',Q'}\in [\mathcal{G}^{(K,Q)}_{\mathcal{I}_{R,J}}]_1}\,\sum_{n=1}^{\infty}\frac{1}{n!}\,ad^{n}Z_{\mathcal{I}^*_{K,Q}}(V^{(K,Q)_{-1}}_{\mathcal{I}_{K',Q'}})\label{V-contr-1}\\
& &+\sum_{\mathcal{I}^*_{K',Q'}\in [\mathcal{G}^{(K,Q)}_{\mathcal{I}_{R,J}}]_2}\,\sum_{n=1}^{\infty}\frac{1}{n!}\,ad^{n}Z_{\mathcal{I}^*_{K,Q}}(V^{(K,Q)_{-1}}_{\overline{\mathcal{I}^*_{K',Q'}}})\label{V-contr-2}\\
& &+\delta_{\widetilde{\mathcal{I}}^*_{K,Q}=\mathcal{I}_{R,J}} \sum_{\mathcal{I}^*_{K',Q'}\in [\mathcal{G}^{(K,Q)}_{\mathcal{I}_{R,J}}]_3}\,\sum_{n=1}^{\infty}\frac{1}{n!}\,ad^{n}Z_{\mathcal{I}^*_{K,Q}}(V^{(K,Q)_{-1}}_{\overline{\mathcal{I}^*_{K',Q'}}})\label{Valgo66}\\
& &{+\delta_{\widetilde{\mathcal{I}}^*_{K,Q}=\mathcal{I}_{R,J}} [P^{(-)}_{\mathcal{I}^*_{K,Q}}V^{(K,Q)_{-1}}_{\mathcal{I}_{K,Q}}P^{(-)}_{\mathcal{I}^*_{K,Q}}-\omega(V^{(K,Q)_{-1}}_{\mathcal{I}_{K,Q}})P^{(-)}_{\mathcal{I}^*_{K,Q}}] } \,\,\quad\quad \label{c-off-diag}\\
& &+\delta_{\widetilde{\mathcal{I}}^*_{K,Q}=\mathcal{I}_{R,J}} [\sum_{m=2}^{\infty}{t^{\frac{(m-1)}{2}}}(V^{(K,Q)_{-1}}_{{\mathcal{I}^*_{K,Q}}})^{\text{diag}}_m]\label{corrW1}\\
& &+\delta_{\widetilde{\mathcal{I}}^*_{K,Q}=\mathcal{I}_{R,J}}\Big(\sum_{n=2}^{\infty}\frac{1}{n!}\,ad^{n}Z_{\mathcal{I}^*_{K,Q}}(\,{\frac{\mathcal{P}^{(2)}_{i^\ast_{-}-1,i^\ast_{-}}}{\sqrt{t}}}+{\frac{\mathcal{P}^{(2)}_{i^\ast_{+},i^\ast_{+}+1}}{\sqrt{t}}})\Big) \label{b-23}\\
&&+\delta_{\widetilde{\mathcal{I}}^*_{K,Q}=\mathcal{I}_{R,J}}[P^{(-)}_{\overline{\mathcal{I}^*_{R,J}}}\Big(adZ_{\mathcal{I}^*_{K,Q}}(\,{\frac{\mathcal{P}^{(2)}_{i^\ast_{-}-1,i^\ast_{-}}}{\sqrt{t}}}+{\frac{\mathcal{P}^{(2)}_{i^\ast_{+},i^\ast_{+}+1}}{\sqrt{t}}})\Big)P^{(+)}_{\overline{\mathcal{I}^*_{R,J}}}+h.c.]\,, \,\,\quad\quad \label{b-232}
\end{eqnarray}
where 
\begin{equation}
 \mathcal{G}^{(K,Q)}_{\mathcal{I}_{R,J}}  :=
\Big\{ \, \mathcal{I}_{K',Q'}\in  \mathfrak{I}_{\text{b.ry}}\,\,\vert \,\,\mathcal{I}_{K',Q'}\cap \mathcal{I}_{K,Q}\neq \emptyset \,\,\text{and} \,\,
\widetilde{\mathcal{I}}^*_{K,Q}\cup \mathcal{I}_{K',Q'}=\mathcal{I}_{R,J}\,\,\Big\} \,.\nonumber
\end{equation}
\begin{rem}\label{rem-c-3}
The terms above are all analogous to the ones in c-1) except for the term in (\ref{c-2-off-2}).  A counterpart of   (\ref{c-2-off-2}) cannot be present in c-1) since it consists of operators where the hooked potentials are supported in intervals ${\mathcal{I}_{K',Q'}} \in \mathfrak{I}_{\text{b.ry}}$, hence such operator cannot contribute to a $V$ term.
\end{rem}
\end{itemize}




\end{enumerate}

\end{defn}

In the next theorem we show that the algorithm described above is consistent with the unitary conjugation 
of the Hamiltonian $K_{\Lambda}^{\,(K,Q)_{-1}}(t)$ generated by the operator $Z_{\mathcal{I}^*_{K,Q}}$.
\begin{thm}\label{consistency}
For the Hamiltonian $K_{\Lambda}^{\,(K,Q)}(t)$, defined iteratively by (\ref{fullHamiltonian})-(\ref{K-3}) and Definition 
\ref{def-interactions-multi} above, the following identity holds
\begin{equation}
K_{\Lambda}^{\,(K,Q)}(t)=e^{Z_{\mathcal{I}^*_{K,Q}}}\,K_{\Lambda}^{\,(K,Q)_{-1}}(t)\,e^{-Z_{\mathcal{I}^*_{K,Q}}}.
\end{equation}
\end{thm}

\noindent
\emph{Proof}

We prove the identity claimed in the statement of the theorem by studying the conjugation of each term 
on the right side of the expression given below 
\begin{eqnarray}
& &e^{Z_{\mathcal{I}^*_{K,Q}}}\,K_{\Lambda}^{\,(K,Q)_{-1}}(t)\,e^{-Z_{\mathcal{I}^*_{K,Q}}}\\
&= &e^{Z_{\mathcal{I}^*_{K,Q}}}\,\Big[\,H_{\Lambda}\\
& &\quad+\sqrt{t}\sum_{Q'}V^{\,(K,Q)_{-1}}_{\overline{\mathcal{I}^*_{1,Q'}}}+\dots+{\sqrt{t}}\sum_{Q'\,;\, (K,Q')\preceq (K,Q)}V^{\,(K,Q)_{-1}}_{\overline{\mathcal{I}^*_{K,Q'}}}\quad\quad\quad \label{Kconj}\\
& &\quad+\sqrt{t}\sum_{Q'\,;\, (K,Q')\succ (K,Q)}V^{\,(K,Q)_{-1}}_{\mathcal{I}_{K,Q'}}+\sqrt{t}\sum_{Q'}
V^{\,(K,Q)_{-1}}_{\mathcal{I}_{K+1,Q'}}+\dots+
\sqrt{t}V^{\,(K,Q)_{-1}}_{\mathcal{I}_{(N-1)\cdot \sqrt{t}-2,2}}\nonumber\\
& &\quad+\sqrt{t}\sum_{Q'}W^{\,(K,Q)_{-1}}_{\mathcal{I}_{1,Q'}}+\dots+
\sqrt{t}W^{\,(K,Q)_{-1}}_{\mathcal{I}_{(N-1)\cdot \sqrt{t},1}}\Big]\,e^{-Z_{\mathcal{I}^*_{K,Q}}}\nonumber
\end{eqnarray}
and subsequently re-assembling the terms according to the rules introduced in Definition \ref{def-interactions-multi}. 

\noindent
The following observations are important.

\begin{enumerate}
\item[(i)]
For all intervals  $\mathcal{I}_{R,J}$ or $\mathcal{I}^*_{R,J}$ with the property  that $\mathcal{I}_{R,J} \cap \mathcal{I}^*_{K,Q}=\emptyset$ or $\mathcal{I}^*_{R,J} \cap \mathcal{I}^*_{K,Q}=\emptyset$, we have that
\begin{eqnarray}
e^{Z_{\mathcal{I}^*_{K,Q}}}\,V^{(K,Q)_{-1}}_{\mathcal{I}_{R,J}}\,e^{-Z_{\mathcal{I}^*_{K,Q}}}&=&V^{(K,Q)_{-1}}_{\mathcal{I}_{R,J}}=:V^{(K,Q)}_{\mathcal{I}_{R,J}}\,,\\
e^{Z_{\mathcal{I}^*_{K,Q}}}\,V^{(K,Q)_{-1}}_{\overline{\mathcal{I}^*_{R,J}}}\,e^{-Z_{\mathcal{I}^*_{K,Q}}}&=&V^{(K,Q)_{-1}}_{\overline{\mathcal{I}^*_{R,J}}}=:V^{(K,Q)}_{\overline{\mathcal{I}^*_{R,J}}}\,,\label{middle}\\
e^{Z_{\mathcal{I}^*_{K,Q}}}\,W^{(K,Q)_{-1}}_{\mathcal{I}_{R,J}}\,e^{-Z_{\mathcal{I}^*_{K,Q}}}&=&W^{(K,Q)_{-1}}_{\mathcal{I}_{R,J}}=:W^{(K,Q)}_{\mathcal{I}^*_{R,J}}\,,
\end{eqnarray}
which follows from a-1) and a-2) in  Definition \ref{def-interactions-multi}.
\item[(ii)]
Using a Lie-Schwinger block-diagonalization associated with an ``unperturbed'' Hamiltonian $G_{\mathcal{I}^*_{K,Q}}$ -- see (\ref{expression-G}) -- and a ``perturbation'' $\sqrt{t}V^{(K,Q)_{-1}}_{\mathcal{I}_{K,Q}}$, we find that
\begin{eqnarray}
& &e^{Z_{\mathcal{I}^*_{K,Q}}}\,\,\Big(H_{\mathcal{I}^*_{K,Q}}^{0}+{\sqrt{t}}\sum_{J=1}^{K-1}\,\,\sum_{\mathcal{I}^*_{J,Q'}\subset \mathcal{I}^*_{K,Q}} V^{(K,Q)_{-1}}_{\overline{\mathcal{I}^*_{J,Q'}}}+\sqrt{t}V^{(K,Q)_{-1}}_{\mathcal{I}_{K,Q}}\Big)\,e^{-Z_{\mathcal{I}^*_{K,Q}}}\,\label{2.71}\\
&= & H_{\mathcal{I}^*_{K,Q}}^{0}+{\sqrt{t}}\sum_{J=1}^{K-1}\,\,\sum_{\overline{\mathcal{I}^*_{J,Q'}}\subset \mathcal{I}^*_{K,Q}} V^{(K,Q)_{-1}}_{\overline{\mathcal{I}^*_{J,Q'}}}+\sqrt{t}\sum_{m=1}^{\infty}{t^{\frac{m-1}{2}}}(V^{(K,Q)_{-1}}_{\mathcal{I}^*_{K,Q}})^{\text{diag}}_m\,\label{lie-sch} \\
&&+e^{Z_{\mathcal{I}^*_{K,Q}}}\sqrt{t}\sum_{\mathcal{I}^*_{J,Q'}\subset \mathcal{I}^*_{K,Q}\, ; \overline{\mathcal{I}^*_{J,Q'}}\nsubset \mathcal{I}^*_{K,Q}}V^{(K,Q)_{-1}}_{\overline{\mathcal{I}^*_{J,Q'}}} e^{-Z_{\mathcal{I}^*_{K,Q}}}\label{touch-terms}
\end{eqnarray}
where in the expression within parentheses in (\ref{2.71}) we have used the identity
\begin{equation}
\sum_{J=1}^{K-1}\,\,\sum_{\mathcal{I}^*_{J,Q'}\subset \mathcal{I}^*_{K,Q}} V^{(K,Q)_{-1}}_{\overline{\mathcal{I}^*_{J,Q'}}}\,=\,\sum_{J=1}^{K-1}\,\,\sum_{\overline{\mathcal{I}^*_{J,Q'}}\subset \mathcal{I}^*_{K,Q}} V^{(K,Q)_{-1}}_{\overline{\mathcal{I}^*_{J,Q'}}}+\sum_{\mathcal{I}^*_{J,Q'}\subset \mathcal{I}^*_{K,Q}\, ; \overline{\mathcal{I}^*_{J,Q'}}\nsubset \mathcal{I}^*_{K,Q}}V^{(K,Q)_{-1}}_{\overline{\mathcal{I}^*_{J,Q'}}}\,,\quad
\end{equation}
and (\ref{lie-sch}) is the result of the Lie-Schwinger conjugation.
Next, we split the conjugation in (\ref{touch-terms}) into the zero order term and the rest, so as to get
\begin{eqnarray}
& &(\ref{2.71})\\
&= & H_{\mathcal{I}^*_{K,Q}}^{0}+{\sqrt{t}\sum_{J=1}^{K-1}\,\,\sum_{\mathcal{I}^*_{J,Q'}\subset \mathcal{I}^*_{K,Q}} V^{(K,Q)}_{\overline{\mathcal{I}^*_{J,Q'}}}+\sqrt{t}((\ref{b1main}))+\sqrt{t}((\ref{expdecayerror}) \text{ or }(\ref{c-off-diag}))+\sqrt{t}((\ref{lshighorder})\text{ or }(\ref{corrW1}))} \nonumber\\
&&+\sqrt{t}((\ref{Valgo67})\text{ or }(\ref{Valgo66}))\,,
\end{eqnarray}
where the alternatives of the type ``$(\ref{expdecayerror}) \text{ or }(\ref{c-off-diag})$" on the right side of the formula above depend on whether the resulting operator is a bulk- or a boundary-potential; furthermore
we have used Definition \ref{def-interactions-multi}, case a-2), which yields the identity
\begin{equation}
\sum_{J=1}^{K-1}\,\,\sum_{\overline{\mathcal{I}^*_{J,Q'}}\subset \mathcal{I}^*_{K,Q}} V^{(K,Q)_{-1}}_{\overline{\mathcal{I}^*_{J,Q'}}}+\sum_{\mathcal{I}^*_{J,Q'}\subset \mathcal{I}^*_{K,Q}\, ; \overline{\mathcal{I}^*_{J,Q'}}\nsubset \mathcal{I}^*_{K,Q}}V^{(K,Q)_{-1}}_{\overline{\mathcal{I}^*_{J,Q'}}} =\sum_{J=1}^{K-1}\,\,\sum_{\mathcal{I}^*_{J,Q'}\subset \mathcal{I}^*_{K,Q}} V^{(K,Q)}_{\overline{\mathcal{I}^*_{J,Q'}}}\,.\quad
\end{equation}
\item[(iii)]
The action of the conjugation on the terms $V^{(K,Q)_{-1}}_{\mathcal{I}_{R,J}}$, with $\mathcal{I}^*_{K,Q} \subset \mathcal{I}_{R,J}$, is
\begin{equation}
e^{Z_{\mathcal{I}^*_{K,Q}}}\,V^{(K,Q)_{-1}}_{\mathcal{I}_{R,J}}\,e^{-Z_{\mathcal{I}^*_{K,Q}}}=(\ref{identity-c}).
\end{equation}
\item[(iv)]
For the conjugation of the terms $V^{(K,Q)_{-1}}_{\mathcal{I}_{R,J}}$, with $ \mathcal{I}^*_{K,Q} \cap \mathcal{I}_{R,J}\neq \emptyset $ and $\mathcal{I}^*_{K,Q} \nsubset \mathcal{I}_{R,J}$, $\mathcal{I}_{R,J}\nsubset \mathcal{I}^*_{K,Q} $ , 
\begin{equation}\label{272}
e^{Z_{\mathcal{I}^*_{K,Q}}}\,V^{(K,Q)_{-1}}_{\mathcal{I}_{R,J}}\,e^{-Z_{\mathcal{I}^*_{K,Q}}}=V^{(K,Q)_{-1}}_{\mathcal{I}_{R,J}}+\sum_{n=1}^{\infty}\frac{1}{n!}\,ad^{n}Z_{\mathcal{I}^*_{K,Q}}(V^{(K,Q)_{-1}}_{\mathcal{I}_{R,J}})\,,
\end{equation}
we notice that the first term on the right side of (\ref{272}) is $V^{(K,Q)}_{\mathcal{I}_{R,J}}$ (see cases a-1) Definition 
\ref{def-interactions-multi}); as for the second term:
\begin{itemize}
\item if $\mathcal{I}_{R',J'} \equiv \mathcal{I}_{R,J}\cup \widetilde{\mathcal{I}}^*_{K,Q}\in\mathfrak{I}_{\text{bulk}}$ 
it contributes to $V^{(K,Q)}_{\mathcal{I}_{R',J'}}$,  according to (\ref{A-map-1-bis});
\item if $\mathcal{I}_{R',J'} \equiv \mathcal{I}_{R,J}\cup \widetilde{\mathcal{I}}^*_{K,Q}\in\mathfrak{I}_{\text{b.ry}}$ 
it contributes to $W^{(K,Q)}_{\mathcal{I}_{R',J'}}$,  according to (\ref{V-contr-1}).
\end{itemize}

\item[(v)]
 Concerning the conjugation of the terms of the type $V^{(K,Q)_{-1}}_{\overline{\mathcal{I}^*_{R,J}}}$,  we notice that they 
appear in  (\ref{Kconj}) only for $(K,Q)_{-1}\succeq (R,J)$. Thus, for $(K,Q)\succ (R,J)$, 
we study the possible situations:
\begin{itemize}
\item if $\mathcal{I}^*_{R,J} \cap \mathcal{I}^*_{K,Q}=\emptyset$ we refer to (\ref{middle}); 
\item  if $\mathcal{I}^*_{R,J} \cap \mathcal{I}^*_{K,Q}\neq \emptyset$ 
\begin{equation}
e^{Z_{\mathcal{I}^*_{K,Q}}} V^{(K,Q)_{-1}}_{\overline{\mathcal{I}^*_{R,J}}}\,e^{-Z_{\mathcal{I}^*_{K,Q}}}=V^{(K,Q)_{-1}}_{\overline{\mathcal{I}^*_{R,J}}}+\sum_{n=1}^{\infty}\frac{1}{n!}\,ad^{n}Z_{\mathcal{I}^*_{K,Q}}(V^{(K,Q)_{-1}}_{\overline{\mathcal{I}^*_{R,J}}})\,,
\end{equation}
where the first term is $V^{(K,Q)}_{\overline{\mathcal{I}^*_{R,J}}}$, by a-2) of Definition \ref{def-interactions-multi};  regarding the second term,  i.e., 
$$\sum_{n=1}^{\infty}\frac{1}{n!}\,ad^{n}Z_{\mathcal{I}^*_{K,Q}}(V^{(K,Q)_{-1}}_{\overline{\mathcal{I}^*_{R,J}}})\,,$$
\begin{itemize}
\item if $\mathcal{I}_{R',J'}\equiv \widetilde{\mathcal{I}}^*_{K,Q}\cup \widetilde{\mathcal{I}}^*_{R,J}\in \mathfrak{I}_{bulk}$, it contributes to $V^{(K,Q)}_{\mathcal{I}_{R',J'}}$ according to (\ref{Valgo3}) of Definition \ref{def-interactions-multi}.
\item  if $\mathcal{I}_{R',J'}\equiv \widetilde{\mathcal{I}}^*_{K,Q}\cup \widetilde{\mathcal{I}}^*_{R,J}\in \mathfrak{I}_{b.ry}$,  it contributes to $W^{(K,Q)}_{\mathcal{I}_{R',J'}}$ according to (\ref{V-contr-2}) of Definition \ref{def-interactions-multi}.
\end{itemize}
\end{itemize}
\item[(vi)] With regard to the terms $W^{(K,Q)_{-1}}_{\mathcal{I}_{R,J}}$, we observe that:
\begin{itemize}
\item  the case $\mathcal{I}_{R,J} \cap \mathcal{I}^*_{K,Q}=\emptyset$ has already been discussed;
\item  if $\mathcal{I}^*_{K,Q} \subset \mathcal{I}_{R,J}$ the expression
\begin{equation}
e^{Z_{\mathcal{I}^*_{K,Q}}}\,W^{(K,Q)_{-1}}_{\mathcal{I}_{R,J}}\,e^{-Z_{\mathcal{I}^*_{K,Q}}}=W^{(K,Q)_{-1}}_{\mathcal{I}_{R,J}}+\sum_{n=1}^{\infty}\frac{1}{n!}\,ad^{n}Z_{\mathcal{I}^*_{K,Q}}(W^{(K,Q)_{-1}}_{\mathcal{I}_{R,J}})\,
\end{equation} 
contributes to $W^{(K,Q)}_{\mathcal{I}^*_{R,J}}$ according to (\ref{c-2-off-0});
\item  if $ \mathcal{I}^*_{K,Q} \cap \mathcal{I}_{R,J}\neq \emptyset $ and $\mathcal{I}^*_{K,Q} \nsubset \mathcal{I}_{R,J}$, $\mathcal{I}_{R,J}\nsubset \mathcal{I}^*_{K,Q} $, in the expression
\begin{equation}
e^{Z_{\mathcal{I}^*_{K,Q}}} W^{(K,Q)_{-1}}_{\mathcal{I}_{R,J}}\,e^{-Z_{\mathcal{I}^*_{K,Q}}}=W^{(K,Q)_{-1}}_{\mathcal{I}_{R,J}}+\sum_{n=1}^{\infty}\frac{1}{n!}\,ad^{n}Z_{\mathcal{I}^*_{K,Q}}(W^{(K,Q)_{-1}}_{\mathcal{I}_{R,J}})
\end{equation}
the first term defines $W^{(K,Q)}_{\mathcal{I}_{R,J}}$, by a-1) and a-2); the other terms, i.e., 
\begin{equation}
\sum_{n=1}^{\infty}\frac{1}{n!}\,ad^{n}Z_{\mathcal{I}^*_{K,Q}}(W^{(K,Q)_{-1}}_{\mathcal{I}_{R,J}})
\end{equation}
contribute to $W^{(K,Q)}_{\mathcal{I}_{R',J'}}$, with 
$\mathcal{I}_{R',J'}\equiv \widetilde{\mathcal{I}}^*_{K,Q} \cup \mathcal{I}_{R,J}$, according to (\ref{c-2-off-2}) in c-2).
 \end{itemize}
\item[(vii)] We finally consider the terms of the unperturbed Hamiltonian $H^0_{\Lambda}$ supported in the intervals of type $(i,i+1)$ which overlap with $\mathcal{I}^*_{K,Q}$ but are not contained in it; for these terms we have:
\begin{eqnarray}
&&e^{Z_{\mathcal{I}^*_{K,Q}}}\,\,(\frac{\mathcal{P}^{(2)}_{i^\ast_{+}, i^\ast_{+}+1}}{\sqrt{t}}+\frac{\mathcal{P}^{(2)}_{i^\ast_{-}-1, i^\ast_{-}}}{\sqrt{t}})\,e^{-Z_{\mathcal{I}^*_{K,Q}}}\nonumber\\
&=&\frac{\mathcal{P}^{(2)}_{i^\ast_{+}, i^\ast_{+}+1}}{\sqrt{t}}+\frac{\mathcal{P}^{(2)}_{i^\ast_{-}-1, i^\ast_{-}}}{\sqrt{t}}+\sum_{n=1}^{\infty}\frac{1}{n!}\,ad^{n}Z_{\mathcal{I}^*_{K,Q}}(\frac{\mathcal{P}^{(2)}_{i^\ast_{+}, i^\ast_{+}+1}}{\sqrt{t}}+\frac{\mathcal{P}^{(2)}_{i^\ast_{-}-1, i^\ast_{-}}}{\sqrt{t}})\nonumber\\
&=&\frac{\mathcal{P}^{(2)}_{i^\ast_{+}, i^\ast_{+}+1}}{\sqrt{t}}+\frac{\mathcal{P}^{(2)}_{i^\ast_{-}-1, i^\ast_{-}}}{\sqrt{t}}+(\ref{b-hop1})+(\ref{b-hop2})+[((\ref{b-21})+(\ref{b-212}))\text{ or } ((\ref{b-23}) +  (\ref{b-232}))] \,,\nonumber
\end{eqnarray}
where the first two terms contribute (once multiplied by $\sqrt{t}$) to $H^0_\Lambda$ and the alternative ``$[((\ref{b-21})+(\ref{b-212}))\text{ or } ((\ref{b-23}) +  (\ref{b-232}))]$'' depends on whether $\widetilde{\mathcal{I}}^*_{K,Q}$ is a bulk- or a boundary-interval.
\end{enumerate}
\hspace{14cm}{\qed}\\
\setcounter{equation}{0}
\section{Operator norms and control of the flow}\label{final-section}
In this section we shall provide proofs of the following claims.

1) The block-diagonalization flow is well defined.

2) It yields quantitative information on the low energy spectrum of $K_{N}(t)$, as stated in the 
Theorem of Section \ref{main}; (see Theorem \ref{final-thm}). 

The main tool used in our proofs is induction in the steps $(K,Q)$ of our block-diagonalization procedure. 
This induction is described in Theorem \ref{th-norms}. The induction hypothesis used to carry out step 
$(K,Q)$ consists of certain norm bounds on the effective interaction potentials appearing in step 
$(K,Q)_{-1}$ and of a lower bound on the spectral gap of the local Hamiltonian $G_{\mathcal{I}^*_{K,Q}}$. 
The induction step consists in showing that the same bounds then hold after step $(K,Q)$. 

In order to make the proof of Theorem \ref{th-norms} a little less heavy, some ingredients of our 
induction step are deferred to Lemma \ref{gap-bound} and Lemma \ref{control-L-S}, where we 
carry out the induction step for some of the quantities appearing in Theorem \ref{th-norms}, and to 
Sections \ref{hooked-pot} and \ref{hooked-proj},  where we estimate norms of so-called ``hooked terms'', 
starting from the norms of the interaction potentials involved in the ``hooking''.
\subsection{Gap estimate}\label{gap}
 In order to prove the lower bound on the spectral gap of the local Hamiltonian in step $(K,Q)_{+1}$, it is sufficient to bound the operator 
\begin{equation}\label{offdiag-def}
P^{(+)}_{\mathcal{I}^*_{(K,Q)_{+1}}}\,(G_{{\mathcal{I}^*_{(K,Q)_{+1}}}}-E_{{\mathcal{I}^*_{(K,Q)_{+1}}}})\,P^{(+)}_{{\mathcal{I}^*_{(K,Q)_{+1}}}}\,
\end{equation}
from below, where  the \emph{local ground-state energy} is defined in (\ref{def-E-bis}). The argument is essentially the same as in \cite{FP}, but with some non-trivial twists caused by
having to deal with macroscopic and microscopic quantities at the same time; see $\mathfrak{I}3$) below. 
We assume that, for all $(K',Q')\,\preceq\, (K,Q)$,
\begin{equation}\label{ind-ass}
\|V^{(K', Q')_{-1}}_{\mathcal{I}_{K',Q'}}\|\,\leq  \frac{t^{\frac{K'-1}{16}}}{(K')^2}\,,\quad  \|V^{(K, Q)}_{\overline{\mathcal{I}^*_{K',Q'}}}\|\,\leq C_{\varepsilon}\cdot \frac{t^{\frac{K'-1}{16}}}{(K')^2}\,, \quad C_\varepsilon:=\left(3+2\cdot \frac{2 A_1}{\varepsilon}\right)\,,
\end{equation}
where $ A_1$ is a universal constant introduced in Lemma \ref{control-L-S}.
The assumptions in (\ref{ind-ass}) above are shown to hold within the proof by induction, in Theorem \ref{th-norms} and Lemma \ref{control-L-S}.
Next, we describe some consequences of these assumptions which will be used later on.

\begin{itemize}
\item[$\mathfrak{I}1$)] For $(K',Q')\, \preceq\, (K,Q)$, $V^{(K,Q)}_{\overline{{\mathcal{I}^*_{K',Q'}}}}$ is a block-diagonalized potential by construction and -- see (\ref{b1main}) -- corresponds to
\begin{eqnarray}\label{L-S-series-bis}
& &V^{(K,Q)}_{\overline{\mathcal{I}^*_{K',Q'}}}\\
&= &\omega(V^{(K',Q')_{-1}}_{\mathcal{I}_{K',Q'}})+P^{(+)}_{\overline{\mathcal{I}^*_{K',Q'}}}\,P^{(+)}_{\mathcal{I}^*_{K',Q'}}\,\,\Big[V^{(K',Q')_{-1}}_{\mathcal{I}_{K',Q'}}-\omega(V^{(K',Q')_{-1}}_{\mathcal{I}_{K',Q'}})\Big]\,P^{(+)}_{\mathcal{I}^*_{K',Q'}}\,P^{(+)}_{\overline{\mathcal{I}^*_{K',Q'}}}\quad\quad \nonumber \\
& &+P^{(+)}_{\overline{\mathcal{I}^*_{K',Q'}}}\,\Big(ad\,Z_{\mathcal{I}^*_{K',Q'}}(\,\frac{\mathcal{P}^{(2)}_{i^*_{-}-1,i^*_{-}}}{\sqrt{t}})\Big)\,P^{(+)}_{\overline{\mathcal{I}^*_{K',Q'}}}
+P^{(+)}_{\overline{\mathcal{I}^*_{K',Q'}}}\,\Big(ad\,Z_{\mathcal{I}^*_{K',Q'}}(\,\frac{\mathcal{P}^{(2)}_{i^*_{+}, i^*_{+}+1}}{\sqrt{t}})\Big)\,P^{(+)}_{\overline{\mathcal{I}^*_{K',Q'}}} \nonumber
\end{eqnarray}

Furthermore, provided $\overline{\mathcal{I}^*_{K',Q'}}\subset \overline{\mathcal{I}^*_{(K,Q)_{+1}}}$, it is block-diagonal w.r.t. to the pair of projections $P^{(-)}_{\overline{\mathcal{I}^*_{(K,Q)_{+1}}}}$, 
$P^{(+)}_{\overline{\mathcal{I}^*_{(K,Q)_{+1}}}}$, thanks to $P^{(+)}_{\overline{\mathcal{I}^*_{K',Q'}}}\,P^{(-)}_{\overline{\mathcal{I}^*_{(K,Q)_{+1}}}}=0$ which follows easily from Definition \ref{def-proj} { and the \textit{frustration free property} of the AKLT model}.

\item[$\mathfrak{I}2$)] 
 Note that, except for the unperturbed Hamiltonian,  $H_{\mathcal{I}^*_{(K,Q)_{+1}}}^{0}$, the general term in (\ref{offdiag-def}) is given by 
\begin{eqnarray}
&& \Big\{P^{(+)}_{\overline{\mathcal{I}^*_{K',Q'}}}\,P^{(+)}_{\mathcal{I}^*_{K',Q'}}\,\,\Big[V^{(K',Q')_{-1}}_{\mathcal{I}_{K',Q'}}-\omega(V^{(K',Q')_{-1}}_{\mathcal{I}_{K',Q'}})\Big]\,P^{(+)}_{\mathcal{I}^*_{K',Q'}}\,P^{(+)}_{\overline{\mathcal{I}^*_{K',Q'}}}\label{uno-bis}\\
 &&+P^{(+)}_{\overline{\mathcal{I}^*_{K',Q'}}}\Big(ad\,Z_{\mathcal{I}^*_{K',Q'}}(\,\frac{\mathcal{P}^{(2)}_{i^*_{-}-1,i^*_{-}}}{\sqrt{t}})\Big)P^{(+)}_{\overline{\mathcal{I}^*_{K',Q'}}}
+P^{(+)}_{\overline{\mathcal{I}^*_{K',Q'}}}\Big(ad\,Z_{\mathcal{I}^*_{K',Q'}}(\,\frac{\mathcal{P}^{(2)}_{i^*_{+}, i^*_{+}+1}}{\sqrt{t}})\Big)P^{(+)}_{\overline{\mathcal{I}^*_{K',Q'}}}\,\Big\}\,\,\quad\quad \label{due-bis}
\end{eqnarray}
since $\omega(V^{(K,Q)}_{\overline{\mathcal{I}^*_{K',Q'}}})=\omega(V^{(K',Q')_{-1}}_{\mathcal{I}_{K',Q'}})$.

We thus focus on $(\ref{uno-bis})+(\ref{due-bis})$ and we observe that (see (\ref{L-S-series-bis}))
\begin{eqnarray}
&& (\ref{uno-bis})+(\ref{due-bis})\\
&=&P^{(+)}_{\overline{\mathcal{I}^*_{K',Q'}}}\,\,(V^{(K,Q)}_{\overline{\mathcal{I}^*_{K',Q'}}}\,-\,\omega(V^{(K',Q')_{-1}}_{\mathcal{I}_{K',Q'}}))\,P^{(+)}_{\overline{\mathcal{I}^*_{K',Q'}}}\,
\end{eqnarray}

Henceforth, making use of the inequality
\begin{equation}
P^{(+)}_{\overline{\mathcal{I}^*_{K',Q'}}}\leq \frac{1}{\varepsilon}\,H^0_{\overline{\mathcal{I}^*_{K',Q'}}}
\end{equation}
(recall that $\varepsilon$ is a lower bound on the spectral gap of $H^0_{\overline{\mathcal{I}^*_{K',Q'}}}\,$; see Theorem \ref{AKTL-gap}), which follows from the definitions of $P^{(+)}_{\overline{\mathcal{I}^*_{K',Q'}}}$ and $H^0_{\overline{\mathcal{I}^*_{K',Q'}}}\,$, we find that
\begin{eqnarray}
& &\pm\{(\ref{uno-bis})+(\ref{due-bis})\}\label{in-est-V}\\
&=&\pm \{P^{(+)}_{\overline{\mathcal{I}^*_{K',Q'}}}\,\,(V^{(K,Q)}_{\overline{\mathcal{I}^*_{K',Q'}}}\,-\,\omega(V^{(K',Q')_{-1}}_{\mathcal{I}_{K',Q'}}))\,P^{(+)}_{\overline{\mathcal{I}^*_{K',Q'}}} \} \label{fin-est-penultimo}\\
&\leq &\frac{2 C_\varepsilon}{\varepsilon}\cdot t^{\frac{K'-1}{16}}\,H^0_{\overline{\mathcal{I}^*_{K',Q'}}}\,, \label{fin-est-V}
\end{eqnarray}
where we have used that $$\|(\ref{fin-est-penultimo})\|\leq \Big[\|V^{(K,Q)}_{\overline{\mathcal{I}^*_{K',Q'}}}\|+\|V^{(K',Q')_{-1}}_{\mathcal{I}_{K',Q'}}\|\Big]\frac{1}{\varepsilon}\,H^0_{\overline{\mathcal{I}^*_{K',Q'}}}\,$$ and the bounds in (\ref{ind-ass}).


\item[$\mathfrak{I}3$)] 
\noindent
Next,  for
$1\leq Q_{\text{min}} \leq Q_{\text{max}} \leq (N-1)\,\sqrt{t}-K'+1$, we set $\mathscr{I}:=\bigcup\limits_{Q'=Q_{\text{min}}}^{Q_{\text{max}}}\,\overline{\mathcal{I}^*_{K',Q'}}$ and observe that
\begin{eqnarray}
\sum_{Q'=Q_{\text{min}}}^{Q_{\text{max}}}H^0_{\overline{\mathcal{I}^*_{K',Q'}}}&=&\sum_{Q'=Q_{min}}^{Q_{max}}\,\sum_{i,i+1\,\in \overline{\mathcal{I}^*_{K',Q'}}}\,\mathcal{P}^{(2)}_{i,i+1}
\label{sum-H-in}\\
&\leq &(K'+1) \sum_{i,i+1\,\in  \mathscr{I}}\, \mathcal{P}^{(2)}_{i,i+1} \,\\
&=&(K'+1)\, H^0_{ \mathscr{I}}\,. \label{sum-H-fin}
\end{eqnarray}

\item[$\mathfrak{I}4$)] 
Using the bound in (\ref{in-est-V})-(\ref{fin-est-V}) and inequalities (\ref{sum-H-in})-(\ref{sum-H-fin}),  we 
conclude that
\begin{eqnarray}
& &\pm\Big\{\sum_{\overline{\mathcal{I}^*_{K',Q'} }\,\subset \, \mathcal{I}^*_{(K,Q)_{+1}}}\,P^{(+)}_{\overline{\mathcal{I}^*_{K',Q'}}}\,P^{(+)}_{\mathcal{I}^*_{K',Q'}}\,\,\Big[V^{(K',Q')_{-1}}_{\mathcal{I}_{K',Q'}}-\omega(V^{(K',Q')_{-1}}_{\mathcal{I}_{K',Q'}})\Big]\,P^{(+)}_{\mathcal{I}^*_{K',Q'}}\,P^{(+)}_{\overline{\mathcal{I}^*_{K',Q'}}}\\
&&\quad +\sum_{\overline{\mathcal{I}^*_{K',Q'}} \, \subset \, \mathcal{I}^*_{(K,Q)_{+1}}}\,P^{(+)}_{\overline{\mathcal{I}^*_{K',Q'}}}\Big(adZ_{\mathcal{I}^*_{K',Q'}}(\,\frac{\mathcal{P}^{(2)}_{i^*_{-}-1,i^*_{-}}}{\sqrt{t}})\Big)P^{(+)}_{\overline{\mathcal{I}^*_{K',Q'}}}
+P^{(+)}_{\overline{\mathcal{I}^*_{K',Q'}}}\Big(adZ_{\mathcal{I}^*_{K',Q'}}(\,\frac{\mathcal{P}^{(2)}_{i^*_{+}, i^*_{+}+1}}{\sqrt{t}})\Big)P^{(+)}_{\overline{\mathcal{I}^*_{K',Q'}}}\,\Big\} \nonumber\\
&\leq &\frac{2C_\varepsilon}{\varepsilon}\cdot t^{\frac{K'-1}{16}}\,(K'+1)\,H^0_{\mathcal{I}^*_{(K,Q)_{+1}}} \,.\label{bound-G}
\end{eqnarray}
\end{itemize}

The items discussed above are the ingredients of the proof of the following result (for more details concerning this proof we refer to {\cite[Section 2.3]{FP}}).

\begin{lem}\label{gap-bound}
Assuming that the bound in (\ref{ind-ass}) holds in step $(K,Q)$ of the block-diagonalization, and choosing $t>0$ so small that
\begin{equation}
\Big\{{ 1}-\frac{4 C_\varepsilon}{\varepsilon}\cdot t^{\frac{1}{2}}-\frac{2 C_\varepsilon}{\varepsilon}\cdot t^{\frac{1}{2}} \sum_{l=3}^{\infty}l\cdot t^{\frac{l-2}{16}}\Big\}>0\,,\label{gap-bound-formula}
\end{equation}
the inequality
\begin{equation}
P^{(+)}_{\mathcal{I}^*_{(K,Q)_{+1}}}\,(G_{{\mathcal{I}^*_{(K,Q)_{+1}}}}-E_{{\mathcal{I}^*_{(K,Q)_{+1}}}})\,P^{(+)}_{{\mathcal{I}^*_{(K,Q)_{+1}}}}
\geq\,\varepsilon\cdot \Big\{1-\frac{4 C_\varepsilon}{\varepsilon}\cdot t^{\frac{1}{2}}-\frac{2 C_\varepsilon}{\varepsilon}\cdot t^{\frac{1}{2}} \sum_{l=3}^{\infty}l\cdot t^{\frac{l-2}{16}}\Big\}\,P^{(+)}_{\mathcal{I}^*_{(K,Q)_{+1}}} \label{final-eq-1}
\end{equation}
holds,  where 
\begin{equation}\label{def-E-bis}
E_{\mathcal{I}^*_{(K,Q)_{+1}}}:=\,{\sqrt{t} }\sum_{J=1}^{K-1}\,\,\sum_{\overline{\mathcal{I}^*_{J,Q'}}\,\subset \,\mathcal{I}^*_{(K,Q)_{+1}}} \,\omega(V^{(K,Q)}_{\overline{\mathcal{I}^*_{J,Q'}}})\,.
\end{equation}.
\end{lem}
\subsection{Preliminary estimates of the operator norms of potentials}
\subsubsection{Estimate of the ``hooked'' potentials}\label{hooked-pot}
Assuming the induction hypothesis (\ref{ind-ass}) and the bounds (\ref{bound-V})-(\ref{bound-S}) proven in Lemma \ref{control-L-S}, we readily conclude that, for sufficiently small $t>0$,
\begin{eqnarray}
& &\Big\|\sum_{n=1}^{\infty}\frac{1}{n!}\,ad^{n}Z_{\mathcal{I}^*_{K,Q}}(V^{(K,Q)_{-1}}_{\overline{\mathcal{I}^*_{K',Q'}}})\Big\|\\
&\leq &C\cdot  A_1\cdot \frac{2}{\varepsilon}\cdot \sqrt{t}\cdot
\|V^{(K,Q)_{-1}}_{\mathcal{I}_{K,Q}}\|\cdot \|V^{(K,Q)_{-1}}_{\overline{\mathcal{I}^*_{K',Q'}}}\|\\
&\leq & C_\varepsilon\cdot C\cdot  A_1\cdot \frac{2}{\varepsilon} \cdot \sqrt{t}\cdot 
\|V^{(K,Q)_{-1}}_{\mathcal{I}_{K,Q}}\|\cdot \|V^{(K',Q')_{-1}}_{\mathcal{I}_{K',Q'}}\|\,,
\end{eqnarray}
where $C$ is a  universal constant and $C_{\varepsilon}$ is defined in (\ref{ind-ass}).
Similarly, we can prove that
\begin{equation}
\Big\|\sum_{n=1}^{\infty}\frac{1}{n!}\,ad^{n}Z_{\mathcal{I}^*_{K,Q}}(V^{(K,Q)_{-1}}_{\mathcal{I}_{K',Q'}})\Big\|\leq C\cdot  A_1\cdot \frac{2}{\varepsilon} \cdot\sqrt{t}\cdot 
\|V^{(K,Q)_{-1}}_{\mathcal{I}_{K,Q}}\|\cdot \|V^{(K,Q)_{-1}}_{\mathcal{I}_{K',Q'}}\|
\end{equation}
and
\begin{equation}
\Big\|\sum_{n=1}^{\infty}\frac{1}{n!}\,ad^{n}Z_{\mathcal{I}^*_{K,Q}}(W^{(K,Q)_{-1}}_{{\mathcal{I}_{K',Q'}}})\Big\|\leq C\cdot  A_1\cdot  \frac{2}{\varepsilon}\cdot \sqrt{t}\cdot \|V^{(K,Q)_{-1}}_{\mathcal{I}_{K,Q}}\|\cdot \|W^{(K,Q)_{-1}}_{{\mathcal{I}_{K',Q'}}}\|\,.
\end{equation}
\subsubsection{Estimate of the off-diagonal part of the hooked projections}\label{hooked-proj}
In this section we assume (\ref{ind-ass}) and  (for $\mathcal{I}_{K,Q}=\mathcal{I}_{R,J}$) the gap bound stated in $\mathcal{S}2$ of Theorem \ref{th-norms} through Lemma \ref{gap-bound}, i.e., 
\begin{equation}\label{gap-bound-formula}
P^{(+)}_{\mathcal{I}^*_{R,J}}(G_{\mathcal{I}^*_{R,J}}-E_{\mathcal{I}^*_{R,J}})P^{(+)}_{\mathcal{I}^*_{R,J}}\geq \frac{\varepsilon}{2}P^{(+)}_{\mathcal{I}^*_{R,J}}\,,
\end{equation} 
then we prove that for $t$ small
\begin{equation}\label{bound-hooking-proj}
\Big\|P^{(+)}_{{\overline{\mathcal{I}^*_{R,J}}}}\,\Big(ad\,Z_{\mathcal{I}^*_{R,J}}(\,{\frac{\mathcal{P}^{(2)}_{i^*_{-}-1, i^*_{-}}}{\sqrt{t}}})\Big)\,P^{(-)}_{{\overline{\mathcal{I}^*_{R,J}}}}\Big\|\leq \mathcal{O}(\frac{t^{\frac{1}{4}}}{\varepsilon^2} \cdot \|V^{(R, J)_{-1}}_{\mathcal{I}_{R,J}}\|)\,.
\end{equation}
From the definition of $ad$, and using $\mathcal{P}^{(2)}_{i^*_{-}-1, i^*_{-}}\,P^{(-)}_{{\overline{\mathcal{I}^*_{R,J}}}}=0$ in the step from (\ref{uno}) to (\ref{due}), we have
\begin{eqnarray}
P^{(+)}_{{\overline{\mathcal{I}^*_{R,J}}}}\,\Big(ad\,Z_{\mathcal{I}^*_{R,J}}(\,{\frac{\mathcal{P}^{(2)}_{i^*_{-}-1, i^*_{-}}}{\sqrt{t}}})\Big)\,P^{(-)}_{{\overline{\mathcal{I}^*_{R,J}}}}
&=&P^{(+)}_{{\overline{\mathcal{I}^*_{R,J}}}}\,\Big[\,Z_{\mathcal{I}^*_{R,J}}\,,\,{\frac{\mathcal{P}^{(2)}_{i^*_{-}-1, i^*_{-}}}{\sqrt{t}}}\Big]\,P^{(-)}_{{\overline{\mathcal{I}^*_{R,J}}}}\label{uno}\\
&=&-P^{(+)}_{{\overline{\mathcal{I}^*_{R,J}}}}\,{\frac{\mathcal{P}^{(2)}_{i^*_{-}-1, i^*_{-}}}{\sqrt{t}}}\,Z_{\mathcal{I}^*_{R,J}}\,P^{(-)}_{{\overline{\mathcal{I}^*_{R,J}}}}\label{due}\\
&=&-\sum_{j=1}^{\infty}{t^{\frac{ j}{2}}}P^{(+)}_{{\overline{\mathcal{I}^*_{R,J}}}}\,{\frac{\mathcal{P}^{(2)}_{i^*_{-}-1, i^*_{-}}}{\sqrt{t}}}\,(Z_{\mathcal{I}^*_{R,J}})_j\,P^{(-)}_{{\overline{\mathcal{I}^*_{R,J}}}}\,.
\end{eqnarray}
The tail,  starting from $j=2$, of the series above, i.e., 
\begin{equation}\label{partial-expression}
-\sum_{j=2}^{\infty}{t^{\frac{ j}{2}}}P^{(+)}_{{\overline{\mathcal{I}^*_{R,J}}}}\,{\frac{\mathcal{P}^{(2)}_{i^*_{-}-1, i^*_{-}}}{\sqrt{t}}}\,(Z_{\mathcal{I}^*_{R,J}})_j\,P^{(-)}_{{\overline{\mathcal{I}^*_{R,J}}}}\,,
\end{equation}
is norm bounded\footnote{This can be actually shown within the proof of Lemma \ref{control-L-S}.} by $\mathcal{O}(t^{1/2}\cdot \frac{\|V^{(R,J)_{-1}}_{\mathcal{I}_{R,J}}\|^2}{\varepsilon^2})$. Henceforth, we can neglect it  since the bound in  (\ref{bound-hooking-proj}) is fulfilled { for the summand in (\ref{partial-expression})} due to the assumption in (\ref{ind-ass}). As for the leading quantity
 \begin{eqnarray}\label{leading-expression}
& &-t^{\frac{ 1}{2}}P^{(+)}_{\overline{\mathcal{I}^*_{R,J}}}\,\frac{\mathcal{P}^{(2)}_{i^*_{-}-1, i^*_{-}}}{\sqrt{t}}\,(Z_{\mathcal{I}^*_{R,J}})_1\,P^{(-)}_{\overline{\mathcal{I}^*_{R,J}}}\\
&=&-P^{(+)}_{\overline{\mathcal{I}^*_{R,J}}}\,\mathcal{P}^{(2)}_{i^*_{-}-1, i^*_{-}}\,\frac{1}{G_{\mathcal{I}^*_{R,J}}-E_{\mathcal{I}^*_{R,J}}}P^{(+)}_{\mathcal{I}^*_{R,J}}\,V^{(R, J)_{-1}}_{\mathcal{I}_{R,J}}\,P^{(-)}_{\mathcal{I}^*_{R,J}}\,P^{(-)}_{\overline{\mathcal{I}^*_{R,J}}}\,,\label{tre}
\end{eqnarray}
 (where (\ref{tre}) follows from (\ref{leading-expression}) by using the definition in (\ref{formula-V_1})) we exploit the resolvent identity
\begin{equation}
\frac{1}{G_{\mathcal{I}^*_{R,J}}-E_{\mathcal{I}^*_{R,J}}}=\frac{1}{G_{\mathcal{I}^*_{R,J}}-E_{\mathcal{I}^*_{R,J}}+i\delta_t }+\frac{i\delta_t}{G_{\mathcal{I}^*_{R,J}}-E_{\mathcal{I}^*_{R,J}}} \,\frac{1}{G_{\mathcal{I}^*_{R,J}}-E_{\mathcal{I}^*_{R,J}}+i\delta_t }\,;
\end{equation}
here $\delta_t$ is set equal to $t^{\frac{1}{4}}$. Next, from the estimate in  (\ref{gap-bound-formula}),  for $t$ sufficiently small, we can write
\begin{eqnarray}
(\ref{tre})&=&-P^{(+)}_{{\overline{\mathcal{I}^*_{R,J}}}}\,\mathcal{P}^{(2)}_{i^*_{-}-1, i^*_{-}}\,\frac{1}{G_{\mathcal{I}^*_{R,J}}-E_{\mathcal{I}^*_{R,J}}+i\delta_t }P^{(+)}_{\mathcal{I}^*_{R,J}}\,V^{(R, J)_{-1}}_{\mathcal{I}_{R,J}}\,P^{(-)}_{\mathcal{I}^*_{R,J}}\,P^{(-)}_{{\overline{\mathcal{I}^*_{R,J}}}} \\
& &+R_1\label{intermediate}
\end{eqnarray}
with $\|R_1\|\leq \mathcal{O}(\frac{\delta_t}{\varepsilon^2} \cdot  \|V^{(R, J)_{-1}}_{\mathcal{I}_{R,J}}\|)$. 
$R_1$ is a remainder term which does not need further treatment since it fulfills the bound in (\ref{bound-hooking-proj}). On the contrary,  the first term requires some further manipulation: namely  we start implementing  a Neumann expansion of $(G_{\mathcal{I}^*_{R,J}}-E_{\mathcal{I}^*_{R,J}}+i\delta_t)^{-1}$ (see  (\ref{expression-G}), (\ref{def-E}), and (\ref{L-S-series-bis}), with obvious adaptation of the indexes, in order to follow the computation below)
\begin{eqnarray}
& &\frac{1}{G_{\mathcal{I}^*_{R,J}}-E_{\mathcal{I}^*_{R,J}}+i\delta_t }P^{(+)}_{\mathcal{I}^*_{R,J}}\\
 &=&\frac{1}{P^{(+)}_{\mathcal{I}^*_{R,J}} (G_{\mathcal{I}^*_{R,J}}-E_{\mathcal{I}^*_{R,J}}+i\delta_t )P^{(+)}_{\mathcal{I}^*_{R,J}}}P^{(+)}_{\mathcal{I}^*_{R,J}} \\
&=&\frac{1}{H^{0}_{\mathcal{I}^*_{R,J}}+i\delta_t }P^{(+)}_{\mathcal{I}^*_{R,J}}\\
& &+\frac{1}{H^{0}_{\mathcal{I}^*_{R,J}}+i{ \delta_t} }\sum_{j=1}^{\infty}P^{(+)}_{\mathcal{I}^*_{R,J}}\times \label{rem}\\
& &\quad \times\Big\{\,\Big(\, -{\sqrt{t} }\sum_{J=1}^{K-1}\,\,\sum_{\overline{\mathcal{I}^*_{K',Q'}}\subset \mathcal{I}^*_{R,J}} P^{(+)}_{\mathcal{I}^*_{R,J}}\,\Big[V^{(R,J)_{-1}}_{\overline{\mathcal{I}^*_{K',Q'}}}-\omega(V^{(K',Q')_{-1}}_{\mathcal{I}_{K',Q'}})\charf\Big]\,P^{(+)}_{\mathcal{I}^*_{R,J}}\,\Big)\frac{1}{H^{0}_{\mathcal{I}^*_{R,J}}+i\delta_t }\Big\}^j\,P^{(+)}_{\mathcal{I}^*_{R,J}}\nonumber
\end{eqnarray}
that we combine with  (\ref{bound-G}) so as to get the bound
\begin{equation}
\|\,(\ref{rem})\,\|\leq \mathcal{O}(\frac{\sqrt{t}}{\varepsilon^2})\,.
\end{equation}
Therefore we can split the expression as follows
\begin{eqnarray}
& &-P^{(+)}_{{\overline{\mathcal{I}^*_{R,J}}}}\,\mathcal{P}^{(2)}_{i^*_{-}-1, i^*_{-}}\,\frac{1}{G_{\mathcal{I}^*_{R,J}}-E_{\mathcal{I}^*_{R,J}}+i\delta_t }P^{(+)}_{\mathcal{I}^*_{R,J}}\,V^{(R, J)_{-1}}_{\mathcal{I}_{R,J}}\,P^{(-)}_{\mathcal{I}^*_{R,J}}\,P^{(-)}_{{\overline{\mathcal{I}^*_{R,J}}}} \\
&=&-P^{(+)}_{{\overline{\mathcal{I}^*_{R,J}}}}\,\mathcal{P}^{(2)}_{i^*_{-}-1, i^*_{-}}\,\,\frac{1}{H^{0}_{\mathcal{I}^*_{R,J}}+i\delta_t }\,P^{(+)}_{\mathcal{I}^*_{R,J}}\,V^{(R, J)_{-1}}_{\mathcal{I}_{R,J}}\,P^{(-)}_{\mathcal{I}^*_{R,J}}\,P^{(-)}_{{\overline{\mathcal{I}^*_{R,J}}}}\label{H0-left}\\
& &+R_2
\end{eqnarray}
where $\|R_2\|\leq \mathcal{O}(\frac{\sqrt{t}}{\varepsilon^2}\cdot \|V^{(R, J)_{-1}}_{\mathcal{I}_{R,J}}\|)$.
Next, we discard $R_2$ and  substitute $P^{(+)}_{\mathcal{I}^*_{R,J}}=\charf -P^{(-)}_{\mathcal{I}^*_{R,J}}$ in (\ref{H0-left}); the latter expression reads
\begin{eqnarray}
& &-P^{(+)}_{{\overline{\mathcal{I}^*_{R,J}}}}\,\mathcal{P}^{(2)}_{i^*_{-}-1, i^*_{-}}\,\frac{1}{H^{0}_{\mathcal{I}^*_{R,J}}+i\delta_t }P^{(+)}_{\mathcal{I}^*_{R,J}}\,V^{(R, J)_{-1}}_{\mathcal{I}_{R,J}}\,P^{(-)}_{\mathcal{I}^*_{R,J}}\,P^{(-)}_{{\overline{\mathcal{I}^*_{R,J}}}} \\
&=&-P^{(+)}_{{\overline{\mathcal{I}^*_{R,J}}}}\,{\mathcal{P}^{(2)}_{i^*_{-}-1, i^*_{-}}}\,\frac{1}{H^{0}_{\mathcal{I}^*_{R,J}}+i\delta_t }\,V^{(R, J)_{-1}}_{\mathcal{I}_{R,J}}\,P^{(-)}_{\mathcal{I}^*_{R,J}}\,P^{(-)}_{{\overline{\mathcal{I}^*_{R,J}}}}\label{leading-1}\\
& &+P^{(+)}_{{\overline{\mathcal{I}^*_{R,J}}}}\,{\mathcal{P}^{(2)}_{i^*_{-}-1, i^*_{-}}}\,\frac{1}{H^{0}_{\mathcal{I}^*_{R,J}}+i\delta_t }P^{(-)}_{\mathcal{I}^*_{R,J}}\,V^{(R, J)_{-1}}_{\mathcal{I}_{R,J}}\,P^{(-)}_{\mathcal{I}^*_{R,J}}\,P^{(-)}_{{\overline{\mathcal{I}^*_{R,J}}}}\,.\label{remainder-2}
\end{eqnarray}
We notice the identity
\begin{equation}
\frac{1}{H^{0}_{\mathcal{I}^*_{R,J}}+i\delta_t }\,{ P^{(-)}_{\mathcal{I}^*_{R,J}}} V^{(R, J)_{-1}}_{\mathcal{I}_{R,J}}\,P^{(-)}_{\mathcal{I}^*_{R,J}}\,P^{(-)}_{{\overline{\mathcal{I}^*_{R,J}}}}=\frac{1}{i\delta_t}P^{(-)}_{\mathcal{I}^*_{R,J}}\,V^{(R, J)_{-1}}_{\mathcal{I}_{R,J}}\,P^{(-)}_{\mathcal{I}^*_{R,J}}\,P^{(-)}_{{\overline{\mathcal{I}^*_{R,J}}}}
\end{equation}
as a consequence of $H^{0}_{\mathcal{I}^*_{R,J}}P^{(-)}_{\mathcal{I}^*_{R,J}}=0$; next, by exploiting (\ref{LTQO}),  we can estimate 
\begin{eqnarray}
P^{(-)}_{\mathcal{I}^*_{R,J}}\,V^{(R, J)_{-1}}_{\mathcal{I}_{R,J}}\,P^{(-)}_{\mathcal{I}^*_{R,J}}\,P^{(-)}_{{\overline{\mathcal{I}^*_{R,J}}}}\,&=&\, \omega( V^{(R, J)_{-1}}_{\mathcal{I}_{R,J}})\, P^{(-)}_{\mathcal{I}^*_{R,J}}\,P^{(-)}_{{\overline{\mathcal{I}^*_{R,J}}}}+R_3\\
&=& \omega(V^{(R, J)_{-1}}_{\mathcal{I}_{R,J}}) \,P^{(-)}_{{\overline{\mathcal{I}^*_{R,J}}}}+R_3
\end{eqnarray}
where $\|R_3\|\leq \mathcal{O}(3^{-\frac{\sqrt{t^{-1}}}{3}}\cdot \|V^{(R, J)_{-1}}_{\mathcal{I}_{R,J}}\|)$ and $\omega(V^{(R, J)_{-1}}_{\mathcal{I}_{R,J}}) $ is defined in (\ref{def-omega}). Hence, since  $\mathcal{P}^{(2)}_{i^*_{-}-1, i^*_{-}}{ P^{(-)}_{\overline{\mathcal{I}^*_{R,J}}}}=0$, we deduce that
\begin{equation}
\|(\ref{remainder-2})\|\leq \mathcal{O}(\frac{1}{\delta_t}\cdot 3^{- \frac{\sqrt{t^{-1}}}{3}}\cdot \|V^{(R, J)_{-1}}_{\mathcal{I}_{R,J}}\|)\leq \mathcal{O}(t^{1/2}\cdot \|V^{(R, J)_{-1}}_{\mathcal{I}_{R,J}}\|)\,.
\end{equation}
The expression in (\ref{leading-1}), i.e., 
\begin{equation}
-P^{(+)}_{{\overline{\mathcal{I}^*_{R,J}}}}\,{\mathcal{P}^{(2)}_{i^*_{-}-1, i^*_{-}}}\,\frac{1}{H^{0}_{\mathcal{I}^*_{R,J}}+i\delta_t }\,V^{(R, J)_{-1}}_{\mathcal{I}_{R,J}}\,P^{(-)}_{\mathcal{I}^*_{R,J}}\,P^{(-)}_{{\overline{\mathcal{I}^*_{R,J}}}}\,,\label{hooking-2}
\end{equation}
is to be controlled now. For this purpose, we make use of
\begin{equation}\label{int}
\frac{1}{H^{0}_{\mathcal{I}^*_{R,J}}+i\delta_t }=-i\,\int_{0}^{t^{-\frac{1}{3}}}\,e^{i(H^{o}_{\mathcal{I}^*_{R,J}}+i\delta_t) s}\,ds-i\,\int_{t^{-\frac{1}{3}}}^{+\infty}\,e^{i(H^{o}_{\mathcal{I}^*_{R,J}}+i\delta_t) s}\,ds
\end{equation}
and define
\begin{equation}
R_4:=i\,P^{(+)}_{\overline{\mathcal{I}^*_{R,J}}}\,\mathcal{P}^{(2)}_{i^*_{-}-1, i^*_{-}}\,\int_{t^{-1/3}}^{+\infty}\,e^{i(H^{0}_{\mathcal{I}^*_{R,J}}+i\delta_t) s}\,ds\,V^{(R, J)_{-1}}_{\mathcal{I}_{R,J}}\,P^{(-)}_{\mathcal{I}^*_{R,J}}\,P^{(-)}_{\overline{\mathcal{I}^*_{R,J}}}
\end{equation}
with $$\|R_4\|\leq \mathcal{O}(\frac{e^{-\delta_t\cdot t^{-\frac{1}{3}}}}{\delta_t}\|V^{(R, J)_{-1}}_{\mathcal{I}_{R,J}}\|)\leq  \mathcal{O}(t^{\frac{1}{4}}\, \|V^{(R, J)_{-1}}_{\mathcal{I}_{R,J}}\|)\,.$$
Then, by using (\ref{int}), we can write
\begin{eqnarray}
& &-P^{(+)}_{\overline{\mathcal{I}^*_{R,J}}}\,\mathcal{P}^{(2)}_{i^*_{-}-1, i^*_{-}}\,\frac{1}{H^{0}_{\mathcal{I}^*_{R,J}}+i\delta_t }\,V^{(R, J)_{-1}}_{\mathcal{I}_{R,J}}\,P^{(-)}_{\mathcal{I}^*_{R,J}}\,P^{(-)}_{\overline{\mathcal{I}^*_{R,J}}}-R_4\label{leading-2}\\
&=&i\, \int_{0}^{t^{-1/3}}\,P^{(+)}_{\overline{\mathcal{I}^*_{R,J}}}\,\mathcal{P}^{(2)}_{i^*_{-}-1, i^*_{-}}\,e^{i(H^{0}_{\mathcal{I}^*_{R,J}}+i\delta_t)s}\,V^{(R, J)_{-1}}_{\mathcal{I}_{R,J}}\,P^{(-)}_{\mathcal{I}^*_{R,J}}\,P^{(-)}_{\overline{\mathcal{I}^*_{R,J}}}\,ds\label{lead-step-1}\\
&=&i\,  \int_{0}^{t^{-1/3}}\,P^{(+)}_{\overline{\mathcal{I}^*_{R,J}}}\,\mathcal{P}^{(2)}_{i^*_{-}-1, i^*_{-}}\,e^{-\delta_t \cdot s}\, e^{i\cdot H^{0}_{\mathcal{I}^*_{R,J}}\cdot s}\,V^{(R, J)_{-1}}_{\mathcal{I}_{R,J}}\,e^{-i\cdot H^{0}_{\mathcal{I}^*_{R,J}}\cdot s}P^{(-)}_{\overline{\mathcal{I}^*_{R,J}}}\,ds\label{lead-step-2}\\
&=&i\, \int_{0}^{t^{-1/3}}\,P^{(+)}_{\overline{\mathcal{I}^*_{R,J}}}\,\,e^{-\delta_t \cdot s}\, \Big[\mathcal{P}^{(2)}_{i^*_{-}-1, i^*_{-}}\,,\,e^{i\cdot H^{0}_{\mathcal{I}^*_{R,J}}\cdot s}\,V^{(R, J)_{-1}}_{\mathcal{I}_{R,J}}\,e^{-i\cdot H^{0}_{\mathcal{I}^*_{R,J}}\cdot s}\Big]\,P^{(-)}_{\overline{\mathcal{I}^*_{R,J}}}\,ds\,,\label{lead-step-3}
\end{eqnarray}
where from from (\ref{lead-step-1}) to (\ref{lead-step-2})  we have used 
$$P^{(-)}_{\mathcal{I}^*_{R,J}}=e^{-i\cdot H^{0}_{\mathcal{I}^*_{R,J}}\cdot s}P^{(-)}_{\mathcal{I}^*_{R,J}}$$
and the frustration-free property of the unperturbed Hamiltonian, which in turn implies that $P^{(-)}_{\mathcal{I}^*_{R,J}}\,P^{(-)}_{\overline{\mathcal{I}^*_{R,J}}}=P^{(-)}_{\overline{\mathcal{I}^*_{R,J}}}$,
and  from (\ref{lead-step-2}) to (\ref{lead-step-3})   we have used
$$  \mathcal{P}^{(2)}_{i^*_{-}-1, i^*_{-}}P^{(-)}_{\overline{\mathcal{I}^*_{R,J}}}=0\,.$$
Our last tool is the Lieb-Robinson bound (\ref{LR-bound}), by which for $t$ sufficiently small we can estimate (recall $\delta_t=t^{-\frac{1}{4}}$), 
\begin{eqnarray}
 \|(\ref{leading-2})\| &\leq & t^{-1/4}\cdot \sup_{0\leq s\leq t^{-1/3}}\Big\|\,\Big[\mathcal{P}^{(2)}_{i^*_{-}-1, i^*_{-}}\,,\,e^{i\cdot H^{0}_{\mathcal{I}^*_{R,J}}\cdot s}\,V^{(R, J)_{-1}}_{\mathcal{I}_{R,J}}\,e^{-i\cdot H^{0}_{\mathcal{I}^*_{R,J}}\cdot s}\Big]\,\Big\| \\
 &\leq  & t^{-1/4}\cdot \,.
\frac{4\, \|\mathcal{P}^{(2)}_{i^*_{-}-1, i^*_{-}}\|\cdot \|V^{(R, J)_{-1}}_{\mathcal{I}_{R,J}}\|\cdot \|{F_0}\|}{C_1}\cdot e^{-[d(i^*_{-}\,,\,\mathcal{I}_{R,J}) - 2\, \|\Phi\|_1\cdot  C_1\cdot  t^{-1/3}]}\\
&\leq &e^{-\frac{\sqrt{t^{-1}}}{4}}\cdot \|V^{(R, J)_{-1}}_{\mathcal{I}_{R,J}}\|
\end{eqnarray}
where $d(i^*_{-}\,,\,\mathcal{I}_{R,J})= \frac{\sqrt{t^{-1}}}{3}$,  and $C_1$, $\|\Phi\|_1$ and {$F_0$} are positive constants defined in Section \ref{L-B bounds}.

\noindent
This concludes the proof of the bound in (\ref{bound-hooking-proj}).

\subsection{Main theorem}
We recall that the first step of the block-diagonalization is associated with the pair $(1,2)$. By definition $(1,2)_{-1}=(0,N)$, moreover the potentials
$V^{(0,N)}_{\mathcal{I}_{R,J}}$, {with $\mathcal{I}_{R,J}\in \mathfrak{I}_{\text{bulk}}$,  and $W^{(0,N)}_{\mathcal{I}_{R,J}}$, with $\mathcal{I}_{R,J}\in \mathfrak{I}_{\text{b.dry}}$, coincide with the operators $V_{\mathcal{I}_{R,J}}$ appearing in the bare Hamiltonian $K_{\Lambda}(t)$ (see (\ref{bare-2})).

\begin{thm}\label{th-norms}
{ There exists $\bar{t}>0$ independent of $N$, such that for all $|t|<\bar{t}$,}  { for any $(\hat{K},\hat{Q})\preceq ((N-1)\cdot \sqrt{t},1)_{-1}$,  the Hamiltonians $G_{\mathcal{I}^*_{\hat{K},\hat{Q}}}$ are well defined}, and
\begin{enumerate}
\item[$\mathcal{S}1)$] for any interval  $\mathcal{I}_{R,J}$, with $R\geq 1$, the following operator norms estimates hold
\begin{enumerate}
\item $\|V^{(\hat{K},\hat{Q})}_{\mathcal{I}_{R,J}}\| \leq\frac{t^{\frac{R-1}{16}}}{R^2}\,$ for $(R,J) \, \succ \, (\hat{K},\hat{Q})$, 
\item  $\|W^{(\hat{K},\hat{Q})}_{\mathcal{I}_{R,J}}\|\leq\frac{t^{\frac{R-1}{16}}}{R^2}\,$,
\end{enumerate}
\item[$\mathcal{S}2)$]let $(\mathcal{I}^*_{\hat{K},\hat{Q}})_{+1}$ be the interval of type $\mathcal{I}^*$ associated with the pair $(\hat{K},\hat{Q})_{+1}$, then  the Hamiltonian $G_{(\mathcal{I}^*_{\hat{K},\hat{Q}})_{+1}}$ has a spectral gap $\Delta_{(\mathcal{I}^*_{\hat{K},\hat{Q}})_{+1}}$ above its ground-state energy  bounded below by $\frac{\varepsilon}{2},$
where  $G_{\mathcal{I}^*_{K,Q}}$ is defined in (\ref{expression-G}) for $K> 1$,  and $G_{\mathcal{I}^*_{1,Q}}:=H^0_{\mathcal{I}^*_{1,Q}}$.
\end{enumerate}
\end{thm}

\noindent
\emph{Proof}

\noindent
The inductive proof  in the pair index $(K,Q)$ is implemented as follows. We consider a fixed $(R,J)$ and we show that $\mathcal{S}1)$ and $\mathcal{S}2)$ hold from $(K,Q)=(0,N)$ up to $(K,Q)= ((N-1)\cdot \sqrt{t},1)_{-1}$. In turn, by assuming that $\mathcal{S}1)$ holds for all $V^{(K',Q')}_{\mathcal{I}_{R,J}}$, $W^{(K',Q')}_{\mathcal{I}_{R,J}}$ with $(K',Q') \prec (K,Q)$ and  $\mathcal{S}2)$ for all $(K',Q') \prec (K,Q)$, the same properties are proven to hold for $V^{(K,Q)}_{\mathcal{I}_{R,J}}$, $W^{(K,Q)}_{\mathcal{I}_{R,J}}$, and for  $G_{(\mathcal{I}^*_{K,Q})_{+1}}$.   Next we invoke Lemma \ref{control-L-S} and rigorously define  $Z_{\mathcal{I}^*_{K,Q}}$ and $K_{\Lambda}^{(K,Q)}$.

\noindent
In order to check that  $\mathcal{S}1)$ and $\mathcal{S}2)$ are verified at the initial step corresponding to $(\hat{K},\hat{Q})= (0,N)$,  we observe since that $\mathcal{S}1)$ can be verified by direct computation,  because 
$$\|V^{(0,N)}_{\mathcal{I}_{1,J}}\|= \|V_{\mathcal{I}_{1,J}}\|\leq1\,\quad,\quad \|W^{(0,N)}_{\mathcal{I}_{1,J}}\|= \|V_{\mathcal{I}_{1,J}}\|\leq1$$
and
$\|V_{\mathcal{I}_{R,J}}^{(0,N)}\|=\|W^{(0,N)}_{\mathcal{I}_{R,J}}\|= \|V_{\mathcal{I}_{R,J}}\|=0$ otherwise;
then $\mathcal{S}1)$ follows. As far as $\mathcal{S}2)$ is concerned, the statement is true given that $(0,N)_{+1}=(1, 2)$ and $G_{\mathcal{I}^*_{1,2}}=H^{(0)}_{\mathcal{I}^*_{1,2}}$.

Within the single induction step, the proof consists of different parts where the allowed interval of $t(\geq 0)$ is progressively reduced. One of these parts is provided by Lemma \ref{control-L-S}. The induction ensures that the same $t-$interval  works for  all steps. 
\\

Concerning $\mathcal{S}1)$, we show the proof for the potentials $V^{(\hat{K},\hat{Q})}_{\mathcal{I}_{R,J}}$; with minor modifications the same result can be proved for the potentials $W^{(\hat{K},\hat{Q})}_{\mathcal{I}_{R,J}}$.

\noindent
\emph{Induction step in the proof of $\mathcal{S}1)$}

\noindent
In order to prove  $\mathcal{S}1)$ in step $(\hat{K},\hat{Q})$, we re-expand  down to $(1,2)$, step by step, i.e., we relate the norm of $V^{(K,Q)}_{\mathcal{I}_{R,J}}$ to the ones of the operators in step $(K,Q)_{-1}$ in terms of which  $V^{(K,Q)}_{\mathcal{I}_{R,J}}$ is expressed according to the algorithm. It is then clear that for most of the steps the norm is preserved, i.e., $\|V^{(K,Q)}_{\mathcal{I}_{R,J}}\|=\|V^{(K,Q)_{-1}}_{\mathcal{I}_{R,J}}\|$,  and only for special steps we have nontrivial relations.

\noindent
We recall that, due to the rules of the algorithm displayed in Definition  \ref{def-interactions-multi}, a potential of the type $V^{(\hat{K},\hat{Q})}_{{ \mathcal{I}_{R,J}}}$ has been defined only for $ (R,J) \succ(\hat{K},\hat{Q})$; henceforth the following constraints hold:  $R> \hat{K}$ or  $R=\hat{K}$ and $J>\hat{Q}$.  In addition, we observe that in view of the prescribed enlargement in Definition \ref{enl-tilde}, the $R$ cannot be equal to $2$.
\\

\noindent
We observe that if $R=1$ the proof is straightforwad by taking into account that $ (1,J) \succ(\hat{K},\hat{Q})$ and by applying a-1) in Definition  \ref{def-interactions-multi} repeatedly, so as to get
\begin{equation}
\|V^{(\hat{K},\hat{Q})}_{\mathcal{I}_{1,J}}\|=\|V^{(0,N)}_{\mathcal{I}_{1,J}}\|=1\,.
\end{equation}

\noindent
\emph{General case ($R\geq 3$)}

\noindent
We study the re-expansion step $(K,Q)$ to $(K,Q)_{-1}$}, by considering various cases with the help of Definition  \ref{def-interactions-multi} (recall that $V^{(\hat{K},\hat{Q})}_{\mathcal{I}_{R,J}}$ is defined for $R\geq \hat{K}$):
\begin{itemize}
\item[1)] in case a-1),   and, similarly, in case c-1) along with  the constraint $i_{+}, i_{-} \notin \widetilde{\mathcal{I}}^*_{K,Q}$ where $i_{+}, i_{-}$ are the endpoints of $\mathcal{I}_{R,J}$,  it turns out that
\begin{equation}\label{cons}
\|V^{(K,Q)}_{\mathcal{I}_{R,J}}\|=\|V^{(K,Q)_{-1}}_{\mathcal{I}_{R,J}}\|\,
\end{equation} 
for which we notice that: in case c-1) only (\ref{identity-c}) contributes thanks to $i^{+}, i_{-} \notin \widetilde{\mathcal{I}}^*_{K,Q}$;  in case a-1) the equality is straightforward.
\item[2-i)]
in case c-1) along with the property that $\widetilde{\mathcal{I}}^*_{K,Q}$  contains one amongst $i_{+}, i_{-}$ (the endpoints of $\mathcal{I}_{R,J}$), the contributions to the re-expansion are given in (\ref{A-map-1-bis}) and (\ref{Valgo3}),  from which we have
\begin{eqnarray}\label{arg}
\|V^{(K,Q)}_{\mathcal{I}_{R,J}}\|&\leq& \|V^{(K,Q)_{-1}}_{\mathcal{I}_{R,J}}\|\\
& &+\sum_{\mathcal{I}_{K',Q'}\in [\mathcal{G}^{(K,Q)}_{\mathcal{I}_{R,J}}]_1}\,\Big\|\sum_{n=1}^{\infty}\frac{1}{n!}\,ad^{n}Z_{\mathcal{I}^*_{K,Q}}(V^{(K,Q)_{-1}}_{\mathcal{I}_{K',Q'}})\Big\|\label{hook-V}\\
& &+\sum_{\mathcal{I}^*_{K',Q'}\in [\mathcal{G}^{(K,Q)}_{\mathcal{I}_{R,J}}]_2}\,\Big\|\sum_{n=1}^{\infty}\frac{1}{n!}\,ad^{n}Z_{\mathcal{I}^*_{K,Q}}(V^{(K,Q)_{-1}}_{\overline{\mathcal{I}^*_{K',Q'}}})\Big\|\label{hook-V^*}
\end{eqnarray}
\item[2-ii)]
in case c-1) along with the property that $\widetilde{\mathcal{I}}^*_{K,Q}$  contains both  $i_{+}$ and  $i_{-}$ (the endpoints of $\mathcal{I}_{R,J}$) the re-expansion consists of terms (\ref{Valgo67}), (\ref{expdecayerror}), (\ref{lshighorder}), (\ref{b-21}), (\ref{b-212}),  from which we have
\begin{eqnarray}\label{arg}
&&\|V^{(K,Q)}_{\mathcal{I}_{R,J}}\|\leq \|V^{(K,Q)_{-1}}_{\mathcal{I}_{R,J}}\|\\
& &+ \|(\ref{Valgo67})\| +\|(\ref{expdecayerror})\|+\| (\ref{lshighorder})\|+\| (\ref{b-21})\|+\| (\ref{b-212})\| \label{tilde-case}\,.
\end{eqnarray}
\end{itemize}
The control of $(\ref{hook-V})+(\ref{hook-V^*})$ relies on  the computations in Section \ref{hooked-pot} together with the assumption of $\mathcal{S}1)$ in step $(K,Q)_{-1}$; hence we can bound as follows
\begin{eqnarray}
(\ref{hook-V})+(\ref{hook-V^*})&\leq & C_\varepsilon\cdot C\cdot  A_1\cdot \sqrt{t} \sum_{K'=R-K-2}^{R-1}\,\|V^{(K,Q)_{-1}}_{\mathcal{I}_{K,Q}}\|\cdot \|V^{(K,Q)_{-1}}_{\mathcal{I}_{K',Q'}}\| \\
&\leq &  C_\varepsilon\cdot C\cdot  A_1\cdot \sqrt{t} \sum_{m=0}^{K-1}\, \frac{t^{\frac{K-1}{16}}}{K^2}\cdot \frac{t^{\frac{R-K+m-3}{16}}}{(R-K-2+m)^2}\\
&= &  C_\varepsilon\cdot C\cdot  A_1\cdot \sqrt{t}\cdot t^{\frac{R-4}{16}} \sum_{m=0}^{K-1}\,  \frac{t^{\frac{m}{16}}}{K^2\cdot (R-K-2+m)^2}\\
&\leq & C'_\varepsilon\cdot t^{\frac{5}{16}}\cdot \frac{t^{\frac{R-1}{16}}}{K^2\cdot (R-K-2)^2}\,,
\end{eqnarray}
 where $ C_\varepsilon$ (see Lemma \ref{control-L-S}) and $C'_\varepsilon$ are constants depending on $\varepsilon$.

We can bound the sum of terms in $(\ref{tilde-case})$ by
\begin{equation}
\|(\ref{tilde-case})\|\leq C''_\varepsilon\cdot t^{\frac{1}{4}}\cdot \frac{t^{\frac{R-3}{16}}}{(R-2)^2}=C''_\varepsilon\cdot t^{\frac{1}{8}}\cdot \frac{t^{\frac{R-1}{16}}}{(R-2)^2} \, .
\end{equation}
 for some $\varepsilon$-dependent constant $C_\varepsilon ''$.

We observe that, at fixed $K$,  the occurrence in 2-i) takes place only twice, whereas the one described  in 2-ii) happens once and only for $K=R-2$. In conclusion, starting from $(\hat{K},\hat{Q})$ and re-expanding back down to level $(0,N)$, the following estimate holds provided  $t$ is sufficiently small  and by using the input $ \|V^{(0,N)}_{\mathcal{I}_{R,J}}\|=0$ for $R>1$:
\begin{eqnarray}
\|V^{(\hat{K},\hat{Q})}_{\mathcal{I}_{R,J}}\|&\leq& \|V^{(0,N)}_{\mathcal{I}_{R,J}}\|\\
& &+\sum_{K=1}^{R-3}2\cdot C'_\varepsilon\cdot t^{\frac{5}{16}}\cdot \frac{t^{\frac{R-1}{16}}}{K^2\cdot (R-K-2)^2}+C_\varepsilon'' \cdot t^{\frac{1}{8}}\cdot \frac{t^{\frac{R-1}{16}}}{(R-2)^2} \label{3.74}\\
&\leq &\frac{t^{\frac{R-1}{16}}}{R^2}\,,\label{3.75}
\end{eqnarray}
where, in the step from (\ref{3.74}) to (\ref{3.75}), we can take advantage 
of the extra-factors  $t^{\frac{5}{16}}$ and $t^{\frac{1}{4}}$.

\noindent
\emph{Induction step in the proof of $\mathcal{S}2)$}

\noindent
By means of $\mathcal{S}1)$ in step $(\hat{K},\hat{Q})$ that we have just proven, and assuming $\mathcal{S}2)$ in step $(\hat{K},\hat{Q})_{-1}$,  the required property is a consequence of Lemma \ref{gap-bound}.
\qed

In the next lemma,  we derive the estimate of the operator norm of the bulk potentials after the block-diagonalization. We recall that, by construction (see the algorithm in Definition \ref{def-interactions-multi}), each block-diagonalized (bulk) potential does not change in the successive steps of the flow.
\begin{lem}\label{control-L-S}
Assume that $t>0$ is sufficiently small,  $\|V^{(K,Q)_{-1}}_{\mathcal{I}_{K,Q}}\| \leq \frac{t^{\frac{K-1}{16}}}{K^2}$, and $\Delta_{\mathcal{I}_{K,Q}}\geq \frac{\varepsilon}{2}$. Then, for arbitrary  $N$, $K\geq 1$, and $Q\geq 1$, the inequalities
\begin{equation}\label{bound-S}
\|Z_{\mathcal{I}^*_{K,Q}}\|\leq A_1\cdot \sqrt{t}\cdot \frac{2}{\varepsilon}\,
\|V^{(K,Q)_{-1}}_{\mathcal{I}_{K,Q}}\| 
\end{equation}
\begin{equation}\label{bound-j}
\sum_{j=2}^{\infty}t^{\frac{j-1}{2}}\,\|(V^{(K,Q)_{-1}}_{\mathcal{I}^*_{K,Q}})^{\text{diag}}_j\| \leq  D_\varepsilon \cdot \sqrt{t}\, \|V^{(K,Q)_{-1}}_{\mathcal{I}_{K,Q}}\|
\end{equation}
\begin{equation}\label{bound-V}
\|V^{(K,Q)}_{\overline{\mathcal{I}^*_{K,Q}}}\|\leq C_{\varepsilon}\, \|V^{(K,Q)_{-1}}_{\mathcal{I}_{K,Q}}\|\,
\end{equation}
hold true for some universal constant $A_1$, for $C_\varepsilon:=3+2\cdot A_1\cdot \frac{2}{\varepsilon}$, and $D_\varepsilon$ an $\varepsilon$-dependent constant.
\end{lem}
\textit{Proof} In order to state the inequalities in  (\ref{bound-S}) and (\ref{bound-j}) we can essentially proceed as in  \cite[Lemma A.3]{FP}. However, here we make the gap (i.e., $\varepsilon$) dependence of our constants more explicit; the sufficiently small $t$ is eventually $\varepsilon$-dependent.  The bound in (\ref{bound-V}) is then obtained from (\ref{L-S-series-bis}) as follows:

\begin{equation}
\|V^{(K,Q)}_{\overline{\mathcal{I}^*_{K,Q}}}\|\leq \|V^{(K,Q)_{-1}}_{\mathcal{I}_{K,Q}}\|\,+2\|V^{(K,Q)_{-1}}_{\mathcal{I}_{K,Q}}\|
+\frac{2}{\sqrt{t}}  \|Z_{\mathcal{I}^*_{K,Q}}\| \nonumber\,
\end{equation}
which we combine with (\ref{bound-S}).
\qed

We can now prove the main result of the paper.
\begin{thm}\label{final-thm}
There exists some $\bar{t} > 0$ independent of $N$ such that, for any coupling constant $t\in \mathbb{R}$ with $\vert t \vert < \bar{t}$, and for all $0< N< \infty$,
\begin{enumerate}
\item[(i)]{ the spectrum of $K_{\Lambda}(t)$ is contained  in two disjoint, $t$-dependent regions, $\sigma^{+}$ and $\sigma^{-}$, 
separated by a uniformly positive gap $\Delta_{\Lambda}(t) \geq \frac{\varepsilon}{4}$}, with $\varepsilon$ 
independent of $N$, as specified in Theorem 1.2; i.e., $E'-E''>\Delta_{\Lambda}(t)$, for all $E'\in \sigma^{+}$ and all $E''\in \sigma^{-} $;
\item[(ii)] for any $d\in \mathbb{N} \cap [1\,,\,\frac{N}{2})$, the eigenspace corresponding to the eigenvalues
contained in $\sigma^{-}$ is four-dimensional;  the gaps between these eigenvalues  coincide with the 
gaps between the eigenvalues of the symmetric matrix
\begin{equation}\label{formula second result - bis}
P^{(-)}_{\Lambda}\,\Big(\,t\sum_{i=1}^{d}\,V_{i,i+1}+\,t\,\sum_{i=N-d}^{N-1}V_{i,i+1}\Big)\,P^{(-)}_{\Lambda}\,,
\end{equation} 
up to corrections bounded by $$|t|\cdot 3^{-(d-1)}\,+\,o(|t|)\,.$$ 
\end{enumerate}
\end{thm}

\noindent
\emph{Proof.} As in the rest of this section, we assume that $t>0$, without loss of generality. By using the results of 
Theorem \ref{th-norms} (combined with  Lemma \ref{control-L-S}), in step $(K,Q)^{{\bf{f}}}:=((N-1)\cdot \sqrt{t},1)_{-1}$, we obtain 
the transformed Hamiltonian
\begin{eqnarray}
K_{\Lambda}^{\,(K,Q)^{{\bf{f}}}}(t)
&= &H^0_{\Lambda}\\
& &+{\sqrt{t} }\sum_{Q'}V^{\,(K,Q)^{{\bf{f}}}}_{\overline{\mathcal{I}^*_{1,Q'}}}+\dots+{\sqrt{t} }\sum_{Q'}V^{\,(K,Q)^{{\bf{f}}}}_{\overline{\mathcal{I}^*_{(N-1)\cdot \sqrt{t}-3,Q'}}}+{\sqrt{t} }V^{\,(K,Q)^{{\bf{f}}}}_{\overline{\mathcal{I}^*_{(N-1)\cdot \sqrt{t}-2,2}}}\\
& &+{\sqrt{t} }\sum_{Q'}W^{\,(K,Q)^{{\bf{f}}}}_{\mathcal{I}_{1,Q'}}+\dots+
\sqrt{t} W^{\,(K,Q)^{{\bf{f}}}}_{\mathcal{I}_{(N-1)\cdot \sqrt{t},1}}
\end{eqnarray}
where all the bulk potentials are block-diagonalized. As a next step, we consider the boundary terms all together, i.e., we define
\begin{equation}
\sqrt{t}\,\mathsf{W}:={\sqrt{t} }\sum_{Q'}W^{\,(K,Q)^{{\bf{f}}}}_{\mathcal{I}_{1,Q'}}+\dots+
{\sqrt{t} }W^{\,(K,Q)^{{\bf{f}}}}_{\mathcal{I}_{(N-1)\cdot \sqrt{t},1}}\,,
\end{equation} whose norm is bounded by $\mathcal{O}(\sqrt{t})$ (due to statement b) in Theorem \ref{th-norms}), and we implement a block-diagonalization step w.r.t. the projections \begin{equation}\label{final-proj}
P^{(-)}_{\mathcal{I}_{(N-1)\cdot \sqrt{t},1}}\equiv P^{(-)}_{\Lambda}\,,\,P^{(+)}_{\mathcal{I}_{(N-1)\cdot \sqrt{t},1}}\equiv P^{(+)}_{\Lambda}
\end{equation}
 associated with the whole chain; in this operation the ``bulk" operator
\begin{equation}
\mathsf{G}:=H^0_{\Lambda}+{\sqrt{t} }\sum_{Q'}V^{\,(K,Q)^{{\bf{f}}}}_{\overline{\mathcal{I}^*_{2,Q'}}}+\dots+{\sqrt{t} }\sum_{Q'}V^{\,(K,Q)^{{\bf{f}}}}_{\overline{\mathcal{I}^*_{(N-1)\cdot \sqrt{t}-3,Q'}}}+{\sqrt{t} }V^{\,(K,Q)^{{\bf{f}}}}_{\overline{\mathcal{I}^*_{(N-1)\cdot \sqrt{t}-2,2}}}
\end{equation}
plays the role of the unperturbed Hamiltonian, and we make use of the result $\mathcal{S}2)$ (see Theorem \ref{th-norms}) in step $((N-1)\cdot \sqrt{t},1)_{-1}$. Upon this standard perturbation, the resulting block-diagonalized Hamiltonian is
\begin{equation}
\tilde{K}_{\Lambda}(t):=\mathsf{G}+\sqrt{t}\,\mathsf{W}'
\end{equation}  
where $\mathsf{W}'$
is expressed in terms of operators $(\mathsf{W})_j$, $(Z)_j$ by means of the formulae from (\ref{S-definition-1}) to (\ref{formula-V_j}), starting from the interaction $(\mathsf{W})_1=\mathsf{W}$, from $\mathsf{G}$,  and from its ground-state energy $\mathsf{E}$.
By standard estimates, $\tilde{K}_{\Lambda}(t)$ enjoys the spectral features described in the statement, as explained below. 

\noindent
i) For the claim concerning the bound $$\Delta_{\Lambda}(t) \geq \frac{\varepsilon}{4}\,,$$ it is enough to consider the argument used  to prove Lemma \ref{gap-bound} by adding  the new operator $\sqrt{t}\,\mathsf{W}'$. 

\noindent
ii) Concerning the $4\times 4$ matrix describing the restriction 
\begin{equation}\label{low-matrix}
(\tilde{K}_{\Lambda}(t)-\mathsf{E})\,:\,P^{(-)}_{\mathcal{I}_{(N-1)\cdot \sqrt{t},1}}\mathcal{H}^{(N)}\,\to\, P^{(-)}_{\mathcal{I}_{(N-1)\cdot \sqrt{t},1}} \mathcal{H}^{(N)}\,,
\end{equation}
we observe that, up to a remainder  bounded in norm by $o(t)$, we can replace the Lie Schwinger series 
$$ (\sqrt{t}\,\mathsf{W}'=)\sum_{j=1}^{\infty}t^{\frac{j}{2}}\,(\mathsf{W})^{diag}_j$$
by the leading term $\sqrt{t}\,(\mathsf{W})^{diag}_1$, since $\|(\mathsf{W})^{diag}_j\|\leq \mathcal{O}(\sqrt{t})$ for $j\geq 2$; here $diag$ stands for the diagonal part w.r.t. the projections in (\ref{final-proj}). Hence we can restrict the study to the matrix elements of the operator
\begin{equation}\label{only-W}
P^{(-)}_{\Lambda}\Big\{{\sqrt{t} }\sum_{Q'}W^{\,(K,Q)^{{\bf{f}}}}_{\mathcal{I}_{1,Q'}}+\dots+
{\sqrt{t} }W^{\,(K,Q)^{{\bf{f}}}}_{\mathcal{I}_{(N-1)\cdot \sqrt{t}},1}\Big\}P^{(-)}_{\Lambda}\,.
\end{equation}
Next we show that  in (\ref{only-W}) the sum of all the terms corresponding to intervals of length $R\geq 2$ is, up to a multiple of the identity operator,  a matrix that can be estimated in norm less than $o(t)$.  This can be explained thinking of the growth processes yielding potentials of type $W^{\,(K,Q)^{{\bf{f}}}}_{\mathcal{I}_{R,Q'}}$. 
First of all we recall that, by construction, for $\cI_{R,J}\in \mathfrak{I}_{\text{b.dry}}$ with $R\geq 2$,
\begin{equation}\label{initial-W}
W_{\cI_{R,J}}^{(0,N)}=0.
\end{equation}
Hence all the potentials $W_{\cI_{R,J}}^{(K,Q)}$, with $R\geq 2$, result from successive growth processes described in c-2) of Definition \ref{def-interactions-multi}. In this respect, notice that all operators from (\ref{c-2-off-2}) down to (\ref{b-232}) are surely supported at a distance larger than say $\sqrt{t^{-1}}/2$ from the boundaries, since $\mathcal{I}_{K,Q}$ belongs to $\mathfrak{I}_{bulk}$ by hypothesis.  Then, taking also (\ref{initial-W}) into account, we can conclude that the operator $W_{\cI_{R,J}}^{(K,Q)}$, for $R\geq 2$, is in fact supported   at distance larger than say $\sqrt{t^{-1}}/2$ from the boundaries. By using the \emph{LTQO} property in (\ref{LTQO}) we conclude that
\begin{equation}
P^{(-)}_{\Lambda}\,W_{\cI_{R,J}}^{(K,Q)}\,P^{(-)}_{\Lambda}=\omega(P^{(-)}_{\mathcal{I}_{R,J}}\,W_{\cI_{R,J}}^{(K,Q)})\,P^{(-)}_{\mathcal{I}_{(N-1)\cdot \sqrt{t},1}}+\Delta W_{\cI_{R,J}}^{(K,Q)}  
\end{equation}
where $$\|\Delta W_{\cI_{R,J}}^{(K,Q)}\|\leq \mathcal{O}(3^{-\frac{\sqrt{t^{-1}}}{2}} \|W_{\cI_{R,J}}^{(K,Q)}\|)\,.$$
It is then clear that, up to a multiple of the identity operator, the matrix in (\ref{only-W}) corresponds to
\begin{equation}
P^{(-)}_{\Lambda}\Big\{{\sqrt{t} }\sum_{Q'}W^{\,(K,Q)^{{\bf{f}}}}_{\mathcal{I}_{1,Q'}}+\sum_{Q'}\Delta W^{\,(K,Q)^{{\bf{f}}}}_{\mathcal{I}_{2,Q'}}+\dots
{\sqrt{t} }\Delta W^{\,(K,Q)^{{\bf{f}}}}_{\mathcal{I}_{(N-1)\cdot \sqrt{t}},1}\Big\}P^{(-)}_{\Lambda}
\end{equation}
where for $t$ sufficiently small
\begin{equation}
\|\sum_{Q'}\Delta W^{\,(K,Q)^{{\bf{f}}}}_{\mathcal{I}_{2,Q'}}+\dots
{\sqrt{t} }\Delta W^{\,(K,Q)^{{\bf{f}}}}_{\mathcal{I}_{(N-1)\cdot \sqrt{t}},1}\|\leq \mathcal{O}(3^{-\frac{\sqrt{t^{-1}}}{2}})
\end{equation}
thanks to statement b) in Theorem \ref{th-norms}. 
By collecting all the error terms, and by using Weyl inequalities for hermitian matrices, we can conclude that the differences between the eigenvalues of the $4\times 4$ matrix corresponding to (\ref{low-matrix}) coincide with the shifts between the eigenvalues of the matrix
\begin{equation}\label{3.93}
P^{(-)}_{\Lambda}\sum_{Q'}W^{\,(K,Q)^{{\bf{f}}}}_{\mathcal{I}_{1,Q'}}\,P^{(-)}_{\Lambda}\,=\,P^{(-)}_{\Lambda}\,{\sqrt{t} }\,\Big(V_{\mathcal{I}_{1,1}}+V_{\mathcal{I}_{1,(N-1)\cdot \sqrt{t}}}\Big)\,P^{(-)}_{\Lambda}\,,
\end{equation}
up to  $o(t)$ corrections. By rewriting $V_{\mathcal{I}_{1,1}}$ and $V_{\mathcal{I}_{1,(N-1)\cdot \sqrt{t}}}$ in terms of the nearest-neighbor interaction terms $V_{i,i+1}$, the r-h-s of (\ref{3.93}) reads
\begin{equation}\label{3.94}
P^{(-)}_{\Lambda}\,\Big(\,t\sum_{i=1}^{i'}\,V_{i,i+1}+\,t\,\sum_{i=i''}^{N-1}V_{i,i+1}\Big)\,P^{(-)}_{\Lambda}
\end{equation}
where $i'=\sqrt{t^{-1}}$ and $i''=N-\sqrt{t^{-1}}$; recall Definition \ref{rects}. Next, we observe that the gaps between the eigenvalues of the matrix in (\ref{3.94}) do not change if we subtract a multiple of the identity matrix, namely 
$$ P^{(-)}_{\Lambda}\,\Big(\,t\sum_{i=d+1}^{i'}\,\omega(V_{i,i+1})+\,t\,\sum_{i=i''}^{N-d-1}\omega(V_{i,i+1})\Big)\,P^{(-)}_{\Lambda}\,,$$
where $d\leq i'-1$, so as to study the matrix
\begin{eqnarray}
& &P^{(-)}_{\Lambda}\,\Big(\,t\sum_{i=1}^{d}\,V_{i,i+1}+\,t\,\sum_{i=N-d}^{N-1}V_{i,i+1}\Big)\,P^{(-)}_{\Lambda}\\
& &+P^{(-)}_{\Lambda}\,\Big(\,t\sum_{i=d+1}^{i'}\,[V_{i,i+1}-\omega(V_{i,i+1})]+\,t\,\sum_{i=i''}^{N-d-1}[V_{i,i+1}-\omega(V_{i,i+1})]\Big)\,P^{(-)}_{\Lambda}\,.
\end{eqnarray}
Using the \emph{LTQO} property in (\ref{LTQO}) once again, we prove the bound
\begin{eqnarray}\label{3.95}
& &\Big\|\,P^{(-)}_{\Lambda}\,\Big(\,t\sum_{i=d+1}^{i'}\,[V_{i,i+1}-\omega(V_{i,i+1})]+\,t\,\sum_{i=i''}^{N-d-1}[V_{i,i+1}-\omega(V_{i,i+1})]\Big)\,P^{(-)}_{\Lambda}\,\Big\| \\
&\leq &\,t\sum_{i=d+1}^{i'}\Big\|\,P^{(-)}_{\Lambda}\,[V_{i,i+1}-\omega(V_{i,i+1})]P^{(-)}_{\Lambda}\,\Big\|+\,t\sum_{i=i''}^{N-d-1}\Big\|\,P^{(-)}_{\Lambda}\,[V_{i,i+1}-\omega(V_{i,i+1})]P^{(-)}_{\Lambda}\,\Big\| \nonumber\\
&\leq&\,2\cdot t\cdot \sum_{i=d+1}^{\infty}3^{-(i-1)}=\,t\cdot 3^{-(d-1)}\,.\nonumber
\end{eqnarray}
One can easily generalize the argument to  the range $d<\frac{N}{2}$ by subtracting a suitable multiple of the 
identity matrix, and finally get the result (ii) in the statement of the theorem as a consequence of Weyl inequalities 
for hermitian matrices. In concrete applications only small values of $d$ are interesting, in particular $d \ll i'$.

\noindent



\begin{thebibliography}{LNT2}




\bibitem[AKLT]{AKLT} I. Affleck, T. Kennedy, E.H. Lieb, H. Tasaki \emph{Rigorous results on valence-bond ground states in antiferromagnets} Physical Review Letters. 59(7): 799-802 (1987).


\bibitem[BN]{BN} S. Bachmann, B. Nachtergaele.  \emph{On gapped phases with a continuous symmetry and boundary operators} J. Stat. Phys. 154(1-2): 91-112  (2014)

\bibitem[BH]{BH} S. Bravyi and M.B. Hastings. \emph{A short proof of stability of topological order under local perturbations} Commun. Math. Physics.  \textbf{307}, 609 (2011).

\bibitem[BHM]{BHM} S. Bravyi, M.B. Hastings and S. Michalakis. \emph{Topological quantum order: stability under local perturbations} J. Math. Phys.  \textbf{51}, 093512, (2010).



\bibitem[DFFR]{DFFR} N. Datta, R. Fernandez, J. Fr\"ohlich, L. Rey-Bellet. \emph{Low-Temperature Phase Diagrams of Quantum Lattice Systems. II. Convergent Perturbation Expansions and Stability in Systems with Infinite Degeneracy} Helvetica Physica Acta 69, 752--820 (1996)

\bibitem[DFPR1]{DFPR1} S. Del Vecchio, J. Fr\"ohlich, A. Pizzo, S. Rossi. \emph{Lie-Schwinger block-diagonalization and gapped quantum chains with unbounded interactions}, Comm. Math. Phys. https://link.springer.com/article/10.1007/s00220-020-03878-y

\bibitem[DFPR2]{DFPR2} S.~Del Vecchio, J.~Fr\"ohlich, A.~Pizzo, S.Rossi. \textit{Lie-Schwinger block-diagonalization and gapped quantum chains: analyticity of the ground-state energy},  J. Funct. Anal. V.  279 Issue 8 \\
https://www.sciencedirect.com/science/article/abs/pii/S0022123620302469

\bibitem[DFPR3]{DFPR3} S. Del Vecchio, J. Fr\"ohlich, A. Pizzo, S. Rossi. {\emph{Local} iterative block-diagonalization of gapped Hamiltonians: a new tool in singular perturbation theory}  J. Math. Phys. 63 (2022) 073503, https://doi.org/10.1063/5.0084552.
 
\bibitem[DFP]{DFP} S. Del Vecchio, J. Fr\"ohlich, A. Pizzo. \emph{Block-diagonalization of infinite-volume lattice Hamiltonians with unbounded interactions}  J. Funct. Anal. V. 284, Issue 1, 1 January 2023, 109734

\bibitem[DFPRa]{DFPRa} S. Del Vecchio, J. Fr\"ohlich, A. Pizzo, A. Ranallo {\emph{Low energy spectrum of the XXZ model coupled to a magnetic field} arxiv preprint}  




\bibitem[FFU]{FFU} R. Fernandez, J. Fr\"ohlich, D. Ueltschi. \emph{Mott Transitions in Lattice Boson Models.} Comm. Math. Phys. 266, 777-795 (2006)

\bibitem[FP]{FP} J. Fr\"ohlich, A. Pizzo. \emph{Lie-Schwinger block-diagonalization and gapped quantum chains} Comm. Math. Phys. 375, 2039-2069 (2020)







\bibitem[KT]{KT} T. Kennedy, H. Tasaki.  \emph{Hidden symmetry breaking and the Haldane phase in S = 1 quantum spin chains}  Comm.. Math. Phys. 147, 431-484 (1992)


\bibitem[LSW]{LSW} M. Lemm, A. W. Sandvik, L. Wang. \emph{Existence of a spectral gap in the Affleck-Kennedy-Lieb-Tasaki model
on the hexagonal lattice.} Phys. Rev. Lett., 124(17) (2020).

\bibitem[LR]{LR} E. H. Lieb, D. W. Robinson. \emph{The finite group velocity of quantum spin systems} Comm. in Math. Phys. 28 (3) 251-257 (1972).

\bibitem[LMY]{LMY} A. Lucia, A. Moon, A. Young. \emph{Stability of the spectral gap and grounds state indistinguishability for a decorated AKLT model.} preprint arXiv:2209.01141

\bibitem[MZ]{MZ} S. Michalakis, J.P. Zwolak  \emph{Stability of frustration-free Hamiltonians} Comm. Math. Phys.  322, 277-302, (2013)


\bibitem[MN]{MN} A. Moon, B. Nachtergaele.  \emph{Stability of Gapped Ground State Phases of Spins and Fermions in One Dimension} J. Math. Phys.  59, 091415 (2018)


\bibitem[NS]{NS} B. Nachtergaele, R. Sims. \emph{Locality Estimates for Quantum Spin Systems}  in New Trends in Mathematical Physics 591-614 (2009). Spinger Netherlands.

\bibitem[NSY1]{NSY1} B. Nachtergaele, R. Sims, A. Young. \emph{Stability of the bulk gap for frustration-free topologically ordered quantum lattice systems} Lett Math Phys 114, 24 (2024) 

\bibitem[NSY2]{NSY2}B. Nachtergaele, R. Sims,  A. Young,  \emph{Quasi-Locality Bounds for Quantum Lattice Systems. Part II. Perturbations of Frustration-Free Spin Models with Gapped Ground States}  Ann. Henri  Poincar\'e 23, 393-511 (2022).

\bibitem[NSY]{NSY} B. Nachtergaele, R. Sims, A. Young.  \emph{Lieb-Robinson bounds, the spectral flow, and stability of the spectral gap for lattice fermion systems}  Mathematical Problems in Quantum Physics, pp.93-115

\bibitem[AYLLN]{AYLLN} H. Abdul-Rahman,  A. Young, A. Lucia, M. Lemm, B. Nachtergaele. \emph{A class of two-dimensional AKLT models with a gap} Analytic Trends in Mathematical Physics 2020, https://doi.org/10.1090/conm/741/14917

\bibitem[O1]{O1} Y. Ogata. \emph{A Z2-Index of Symmetry Protected Topological Phases with Time Reversal Symmetry for Quantum Spin Chains.} Comm. Math. Phys., 374 (2):705-734, July 2019.

\bibitem[O2]{O2} Y. Ogata. \emph{A H3(G, T)-valued index of symmetry protected topological phases with on-site finite group symmetry for two-dimensional quantum spin systems.} 2021. arXiv:2101.00426.

\bibitem[O3]{O3} Y. Ogata. \emph{A Z2-index of symmetry protected topological phases with reflection symmetry for quantum spin chains.} Comm. in Math. Phys., 385:1245-1272, 2021.


\bibitem[PW1]{PW1} N. Pomata, T.-C. Wei \emph{Demonstrating the Affleck-Kennedy-Lieb-Tasaki spectral gap on 2D degree-3 lattices.} Phys. Rev. Lett., 124(17), April 2020.


\bibitem[PW2]{PW2}
 N. Pomata, T.-C. Wei. \emph{AKLT models on decorated square lattices are gapped.} Phys. Rev. B, 100:094429, 2019.
 


\bibitem[Y]{Y} D.A. Yarotsky.  \emph{Ground States in Relatively Bounded Quantum Perturbations of Classical Systems} Comm. Math. Phys. 261, 799-819 (2006)

\end{thebibliography}
\end{document}